\documentclass[aps,prd,nofootinbib,twocolumn,superscriptaddress,preprintnumbers,balancelastpage,longbibliography]{revtex4-1}

\usepackage{placeins}
\usepackage{amsmath,amssymb,mathtools,bm}
\usepackage{lipsum}
\usepackage{comment}
\usepackage{gensymb}
\usepackage{physics}
\usepackage{dsfont}
\usepackage{cancel}

\usepackage{tablefootnote}
\usepackage{graphicx, color, hepunits}
\usepackage[dvipsnames]{xcolor}
\usepackage{float}
\usepackage{filecontents}
\usepackage{multirow}
\usepackage{slashed}
\usepackage{pifont}

\usepackage[colorlinks=true 
,urlcolor=purple 
,anchorcolor=blue
,citecolor=blue 
,filecolor=magenta  
,linkcolor=blue 
,menucolor=blue
,linktocpage=true
,pdfproducer=medialab
,pdfa=true
]{hyperref}

\usepackage[utf8]{inputenc}
\usepackage[english]{babel}
\usepackage{soul}
\usepackage{hyperref}

\newcommand{\es}[2] {\begin{equation} \label{#1} \begin{split} #2 \end{split} \end{equation}}
\newcommand{\bea}{\begin{eqnarray}}
\newcommand{\eea}{\end{eqnarray}}

\restylefloat{table}

\begin{document}

\title{Clearing up the Strong $CP$   problem}

\author{Joshua N. Benabou}
\affiliation{Theoretical Physics Group, Lawrence Berkeley National Laboratory, Berkeley, CA 94720, U.S.A.}
\affiliation{Leinweber Institute for Theoretical Physics, University of California, Berkeley, CA 94720, U.S.A.}

\author{Anson Hook}
\affiliation{Maryland Center for Fundamental Physics, Department of Physics, University of Maryland, College
Park, MD 20742, U.S.A.}

\author{Claudio Andrea Manzari}
\affiliation{School of Natural Sciences, Institute for Advanced Study, Princeton, NJ 08540, USA}

\author{Hitoshi Murayama}
\affiliation{Leinweber Institute for Theoretical Physics, University of California, Berkeley, CA 94720, U.S.A.}
\affiliation{Kavli Institute for the Physics and Mathematics of the Universe (WPI), University of Tokyo
Institutes for Advanced Study, University of Tokyo, Kashiwa 277-8583, Japan}
\affiliation{Theoretical Physics Group, Lawrence Berkeley National Laboratory, Berkeley, CA 94720, U.S.A.}

\author{Benjamin R. Safdi}
\affiliation{Theoretical Physics Group, Lawrence Berkeley National Laboratory, Berkeley, CA 94720, U.S.A.}
\affiliation{Leinweber Institute for Theoretical Physics, University of California, Berkeley, CA 94720, U.S.A.}
\affiliation{Theoretical Physics Department, CERN, 1211 Geneva, Switzerland}

\date{\today}

\begin{abstract}
The absence of a neutron electric dipole moment (EDM) constrains the quantum chromodynamics (QCD) theta angle to be less than one part in ten billion, posing the Strong {\textit CP} problem. We revisit two classes of proposed solutions. First, we show that when $P$ or $CP$ is realized as a {\it gauged} discrete symmetry---as can arise in quantum gravity---the vacuum necessarily preserves $CP$, contrary to recent claims that discrete-symmetry solutions fail. Gauged discrete models face model-building challenges, such as avoiding contributions to the neutron EDM after spontaneous $P$ or $CP$ breaking, but in principle have no fundamental obstructions. Second, we critically examine recent arguments that the Strong $CP$ problem is illusory, demonstrating that a nonzero neutron EDM at finite $\bar\theta$ follows directly from well-understood QCD dynamics. Taken together, our results reinforce the reality of the Strong $CP$ problem and highlight gauged discrete-symmetry realizations of $P$ or $CP$ as plausible solutions.
\end{abstract}
\maketitle

\section{Introduction}
\label{sec:intro}

The Strong {\it CP} problem provides one of the most promising avenues at present for learning about physics beyond the Standard Model (SM). The problem is experimentally manifest by the absence of a neutron electric dipole moment (EDM), which constrains the $\bar \theta$ parameter of quantum chromodynamics (QCD), which we describe in detail below, to be $|\bar \theta| \lesssim 10^{-10}$~\cite{Abel:2020pzs} (see~\cite{Kim:2008hd,Hook:2018dlk,Safdi:2022xkm,Bonanno:2025wcv} for reviews).  Given that a priori $\bar \theta$ could take values between $-\pi$ and $\pi$ and no additional symmetry of the SM is restored at the point $ \bar \theta = 0$, this represents a fine-tuning problem.  Moreover, this fine-tuning problem does not appear to have an anthropic explanation~\cite{Ubaldi:2008nf,Dine:2018glh}, which strongly motivates mechanisms that set the neutron EDM to zero. Broadly speaking, there are two classes of solutions to the Strong {\it CP} problem that evade fine tuning: (i) the axion, which is a new light degree of freedom that dynamically sets the neutron EDM to zero~\cite{Peccei:1977hh,Peccei:1977ur,Weinberg:1977ma,Wilczek:1977pj}; and (ii), symmetry-based solutions that use spontaneously-broken $P$ or $CP$ symmetry in the ultraviolet (UV) to keep $\bar \theta \sim 0$ in the infrared (IR) while also generating the $CP$ violation observed in the Cabibbo–Kobayashi–Maskawa (CKM) matrix~\cite{Nelson:1983zb,Barr:1984qx,Babu:1988mw,Babu:1989rb,Barr:1991qx,Lavoura:1997pq,Vecchi:2014hpa,Dine:2015jga,Hall:2018let,Dunsky:2019api,Craig:2020bnv,Bonnefoy:2023afx,Hall:2024qqe,Feruglio:2024ytl,Bonnefoy:2025rvo}.
The second class of solutions benefit from the fact that $\bar \theta$ evolves very mildly in the SM under the renormalization group (RG)~\cite{Ellis:1978hq}.  In both scenarios the Strong {\it CP} problem is associated with new physics, though only in case (i) does the new physics need to be light enough to be observable. 

A number of works have recently cast doubts on the Strong {\it CP} problem and its possible solutions.  Refs.~\cite{Ai:2020ptm,Nakamura:2021meh,Ai:2022htq,Ai:2024vfa,Ai:2024cnp,Schierholz:2024var,Schierholz:2025tns}, for example, argue that the neutron EDM is zero even for $\bar \theta \neq 0$, such that there is no Strong {\it CP} problem whatsoever. These authors claim that traditional calculations suggesting a $\bar \theta$-dependent neutron EDM are mirages arising from an incorrect order of limits in the process of taking the infinite volume limit.  Conversely, Refs.~\cite{Dvali:2005zk,Dvali:2007iv,Dvali:2022fdv,Kaplan:2025bgy} argue that the Strong {\it CP} problem does exist but can only be solved by axions or related physics, as discrete symmetries can remove $CP$-violating terms from the Lagrangian or Hamiltonian but are not enough  to guarantee that the QCD vacuum preserves $CP$. 

In this work we attempt to clear up the above confusion about the Strong {\it CP} problem. We argue that the Strong {\it CP} problem does exist and that it may be solved by 
discrete symmetries, though we argue that gauged $P$ /$CP$ are the most motivated ultraviolet (UV) discrete-symmetry-based solutions. This is because, as we show, gauged $P$ or $CP$ in the UV enforces the vacuum to be $CP$ preserving, while also ensuring no $P$ or $CP$ violating operators in the Lagrangian. The gauged discrete symmetry must be spontaneously broken in the infrared (IR), but all contributions to the neutron EDM are calculable from the dynamics responsible for spontaneous $P$ / $CP$ breaking; in particular, there is no dependence of the neutron EDM on an otherwise unknowable superselection parameter for the QCD vacuum.  
Theories of gauged $P$ or $CP$ emerge naturally in the context of quantum gravity theories that involve compactified extra dimensions. The discrete symmetries arise below the compactification scale as the unbroken subgroups of the continuous, identity-connected Lorentz transformations of the higher-dimensional theory, in some cases in combination with unbroken internal symmetries.  We illustrate this idea through a simple 5D toy model of QCD that leads to gauged $P$ in 4D, and we also illustrate how gauged $P$ and $CP$ may arise in string theory.

The rest of this work is organized as follows. In Sec.~\ref{sec:def} we review the traditional approaches to describing the Strong {\it CP} problem, theta vacua, and computing the neutron EDM. Then, in Sec.~\ref{sec:symmetry_solns_strongCP} we show that to solve the Strong $CP$ problem, $P$/$CP$ have to be symmetries of all quantum correlation functions and not simply at the Lagrangian level, highlighting the case of gauged $P$/$CP$. 
We illustrate how gauged $P$ can arise in the context of gauged Lorentz symmetry of a higher-dimensional theory through a simple 5D example in Sec.~\ref{sec:gauging_CP_QG}, and we discuss how gauged $P$ / $CP$ arise from the basic principle illustrated in the toy example in the context of more UV complete string theory constructions.  In Sec.~\ref{sec:is_strong_cp} we show definitively, for example using the Witten-Veneziano relation, that the Strong {\it CP} problem exists, and we point out the flaws in recent works stating that there is no such problem.  Lastly, in Sec.~\ref{sec:discussion} we speculate that the entire discussion of {\it e.g.} Ref.~\cite{Kaplan:2025bgy} about discrete symmetries not solving the Strong {\it CP} problem is in the swampland. We argue that in theories that emerge from quantum gravity the neutron EDM is always dynamically determined and does not depend on an otherwise unknowable superselection sector parameter. This is in part because quantum gravity theories are conjectured to gauge the Chern–Weil global symmetry associated with ${\rm tr} F \wedge F$~\cite{Heidenreich:2020pkc}, with $F$ the non-abelian field strength, for example through an axion-like-particle or $p$-form field, and gauging this current yields the QCD superselection parameter unobservable through gauge redundancy.

We include a number of Appendices that provide additional insight into the problem of gauged $P$ / $CP$ and, more broadly, the Strong {\it CP} problem. In Appendix.~\ref{app:1+1QED} we work through 1+1D quantum electrodynamics (QED), which is an exactly solvable theory that has a $\theta$ term. We discuss the interpretation of theta as a boundary condition and as a Lagrangian parameter in this theory, and show how theta may emerge from an extra dimensional UV completion in analogy with the Aharonov-Bohm effect. We also highlight that this theory does indeed exhibit a Strong {\it CP} problem in that $\theta$ contributes to physical observables analogous to the neutron EDM, and we show that this problem is solved by gauging $P$.
Appendix~\ref{app:typeI} reviews Type I string theory, where worldsheet parity is gauged; we draw analogies between this well-understood theory and the theories of gauged $P$ / $CP$ in 4D that we discuss in the main body of this work.
Appendices~\ref{app:WV_derivation}, ~\ref{app:CDP}, ~\ref{sec:gauge_non_normal_subgroups} provide additional technical calculations that are referenced in the main work.    

\section{Definition of the problem}
\label{sec:def}

The \emph{Strong CP Problem} refers to the absence of a satisfactory theoretical explanation for why $P$ and $CP$ appear to be extremely well conserved in the strong interactions. In order to address this question, it is useful to first identify the possible sources of $P$/$CP$ violation in QCD that could induce a non-vanishing neutron EDM. The QCD Lagrangian contains a single dim-4 $CP$-violating operator, namely
\begin{equation}
\mathcal{L}_{\rm QCD} \supset -\frac{\theta_L}{32 \pi^2} \, G_{\mu\nu}^a \, \tilde{G}^{a\,\mu\nu} \,,
\label{eq:thetaL}
\end{equation}
where $\theta_L$ is a dimensionless parameter taking values between $0$ and $2 \pi$, $G_{\mu\nu}^a$ is the gluon field strength tensor, and $\tilde{G}^{a\,\mu\nu} = \frac{1}{2}\epsilon^{\mu\nu\rho\sigma}G_{\rho\sigma}^a$ its dual.
Additionally, through a chiral rotation of the quark fields, which is anomalous, we may remove a phase in the quark mass matrix to form the physical parameter combination $\bar \theta \equiv \theta_L - {\rm arg} \, {\rm det}(Y_u Y_d)$ in~\eqref{eq:thetaL}, where $Y_u$ and $Y_d$ are the up and down-type Yukawa matrices, respectively.

The theta term in QCD affects expectation values $\mathcal{O}$ in a straightforward way:
\begin{align}
    \langle \mathcal{O} \rangle = \frac{\sum_{\nu}e^{i(\bar \theta + \theta)\nu}\int_{\nu}[dA]\exp(iS[A])\mathcal{O}[A]}{\sum_{\nu}e^{i(\bar \theta + \theta)\nu}\int_{\nu}[dA]\exp(iS[A])}\,,
    \label{eq:PathIntegral}
\end{align}
where we sum over gauge configurations labeled by their instanton number
\es{}{
    \nu = \frac{1}{8\pi^2}\int {\rm tr} \left( G \wedge G \right) &= {1 \over 16 \pi^2} \int d^4x {\rm tr} \left(G_{\mu \nu} \tilde G^{\mu \nu} \right) \\
    &= {1 \over 32 \pi^2} \int d^4x\, G_{\mu\nu}^a \tilde{G}^{a \mu\nu}\,.
}
Note that in writing~\eqref{eq:PathIntegral} we do not include the theta term in the action $S[A]$, as its contribution is accounted for in the exponential prefactors.
The parameter $\theta$ appearing in~\eqref{eq:PathIntegral} is a free and otherwise undetermined parameter labeling the superselection sector of the theory. The $\theta$ parameter is commonly associated to the theta vacua of the theory, but it may also be interpreted as the choice of boundary condition for the QCD wavefunctional under large gauge transformations (see App.~\ref{app:1+1QED}).  However, $\theta$ and $\bar \theta$ always appear in the linear combination $\bar \theta + \theta$, so in the path integral formulation of quantum field theory it is unphysical and unnecessary to distinguish them.  Still, in the next section we keep these terms distinct to make contact with the claims of Refs.~\cite{Dvali:2005zk,Dvali:2007iv,Dvali:2022fdv,Kaplan:2025bgy}, which we discuss in the next section.

We emphasize that $\theta$ is part of the definition of the theory.  Distinct values of $\theta$ correspond to different theories.  One might wonder, for example, whether the Universe could contain causally disconnected regions characterized by different values of $\theta$; in the Standard Model, the answer is no. If one extends the Standard Model by the inclusion of {\it e.g.} an axion, then there can be an effective $\bar \theta + \theta$ that varies on superhorizon scales, but without a new degree of freedom like an axion this is not possible. 

The $\bar \theta$ dependence of the neutron EDM may then be computed through a number of techniques.\footnote{Here, we absorb $\theta$ into the definition of $\bar \theta$.} The most straightforward, textbook approach is to use chiral perturbation theory ($\chi$-PT), which returns the estimate (see, {\it e.g.},~\cite{Srednicki:2007qs,Hook:2018dlk,Safdi:2022xkm}) 
\es{dn:chipt}{
\left.d_n \right|_{ \chi {\rm PT}} \approx 3 \times 10^{-16} {\bar \theta} \, {\rm e} \cdot {\rm cm} \,,
}
though within $\chi$-PT it is difficult to estimate the uncertainties on this prediction.  A more precise calculation may be done using QCD sum rules~\cite{Pospelov:1999ha}, yielding
\es{dn:sum}{
\left. d_n \right|_{ {\rm QCD \,\, sum  \,\, rule}} \approx 2.4(7) \times 10^{-16} {\bar \theta} \, {\rm e} \cdot {\rm cm} \,.
}
The $\bar \theta$ dependence of the neutron EDM has also recently been computed on the lattice, with the most recent estimate~\cite{Liang:2023jfj}
\es{dn:lattice}{
\left. d_n \right|_{ \rm{lattice \, \, QCD}} \approx 1.48 (14)(31) \times 10^{-16} {\bar \theta} \, {\rm e} \cdot {\rm cm} \,,
}
with the uncertainties statistical and systematic, respectively.  The best measurements of $d_n$ set the bound~\cite{Baker:2006ts,Pendlebury:2015lrz,Graner:2016ses,Abel:2020pzs}
\es{}{
|d_n| \lesssim 1.8 \times 10^{-26} \, \, {\rm e} \cdot {\rm cm} \,,
}
at 90\% confidence, which constraints $|\bar \theta| \lesssim 10^{-10}$.

Thus, it is clear that $\bar \theta \sim 0$ to very good accuracy in nature. On the other hand, why should this be the case, given that the SM violates $P$ and CP?   This is the Strong $CP$ problem. Of course, it is not so much a problem as a hint of possible new dynamics. 
In the next section we address recent claims~\cite{Dvali:2005zk,Dvali:2007iv,Dvali:2022fdv,Kaplan:2025bgy} that discrete symmetries in the UV cannot be used as a mechanism to explain to $\bar \theta \sim 0$.  Following this discussion, in Sec.~\ref{sec:is_strong_cp} we address the even more provocative claims that $d_n$ is actually zero in the SM for any $\bar \theta$, in contrast with the calculation results quoted in~\eqref{dn:chipt},~\eqref{dn:sum}, and~\eqref{dn:lattice}.

\section{Symmetry Solutions to the Strong $CP$ Problem}
\label{sec:symmetry_solns_strongCP}

In the previous section we argue that $\bar \theta$ and $\theta$, the parameter characterizing the superselection sector of the theory, always show up in the combination $\bar \theta + \theta$, so that we may absorb $\theta$ into the definition of $\bar \theta$. While this is the standard story,
Refs.~\cite{Dvali:2005zk,Dvali:2007iv,Dvali:2022fdv,Kaplan:2025bgy} claim that $P$ or $CP$ can in principle be used to remove the $\bar \theta$ term from the Lagrangian, since this term is odd under $P$/$CP$, but cannot be used to remove $\theta$. Let us now discuss why their interpretation is overly simplistic. First, however, we must define what we mean by a discrete symmetry. We give three definitions, following from the usual description of symmetries in quantum field theory:
\begin{enumerate}
\item A classical, discrete, global symmetry---this is a symmetry of the classical equations of motion.
\item A quantum, discrete, global symmetry---this is a non-anomalous symmetry of all quantum correlation functions.
\item A gauged discrete symmetry---this is a local identification of configurations related by the symmetry action.
\end{enumerate}
The requirement that $P$/$CP$ is the first type of symmetry, that is a classical one, is clearly not sufficient to solve the Strong {\it CP} Problem, since the terms involving $\theta$ and $\bar \theta$ in the partition function do not enter into the classical equations of motion.\footnote{Note that throughout the rest of this section we discuss $P$ for simplicity and not $CP$, since we focus purely on the gluon sector without fermions in this section. We comment specifically on $CP$ later in this work, though the general statements of this section apply to both $P$ and $CP$ symmetry solutions.} That is, even at $\theta + \bar \theta \neq 0$, one would conclude that $P$ is a classical symmetry. However, expectation values of operators would not obey $P$ because in this case $P$ is anomalous. That is, the vacuum expectation value of parity odd operators would not need to vanish in this case.  Here, we define an anomalous symmetry to be a symmetry of the classical equations of motion that is broken by quantum effects; this is, a symmetry of the classical but not quantum theory. 

Now, let us suppose that $P$ is a non-anomalous global symmetry (option 2 above). We claim that this is only possible if $\theta + \bar \theta = 0$ or $\pi$.  For $P$ to be an exact symmetry of the quantum theory, the partition function should be invariant under orientation reversals of the underlying manifold. Under such an orientation reversal, the instanton number $\nu$ changes sign: $\nu \to - \nu$.  Since the instanton number enters into the partition function as $e^{i (\theta + \bar \theta) \nu}$, for the theory to be invariant under orientation reversals we need $e^{i (\theta + \bar \theta) \nu} = e^{-i (\theta + \bar \theta) \nu}$. Equivalently, $e^{2 i (\theta + \bar \theta) \nu} = 1$, which implies that $\theta + \bar \theta = 0$ or $\pi$, since $\nu$ is an integer. 

Let us introduce a $\mathbb{Z}_2$-valued classical background field $a$ that is a function of spacetime that implements local orientation reversals.\footnote{To be more precise,  we can introduce a background gauge field as a holonomy, {\it i.e.}, a map $W: C \to \mathbb{Z}_2 $ between closed loops $C$ in spacetime and the $\mathbb{Z}_2$, such that a particle traversing $C$ will return to the starting point with the same (opposite) handedness if $W(C)=1$ ($W(C)=-1$). Alternatively, for a manifold covered by charts $U_i$, each with a choice of coordinates $x_i^\nu$, we can define a background field  as the transition function between overlaps 
$a_{i \rightarrow j}=\operatorname{sign} \operatorname{det}\left(\frac{\partial x_j^\nu}{\partial x_i^\mu}\right) \in \mathbb{Z}_2$. This means that $a_{i \rightarrow j}=1$ on overlaps between charts which have the same orientation and $-1$ where the orientations differ. For a non-orientable manifold such as a M\"obius strip or a Klein bottle, any choice of charts will have some overlap with $a_{i \rightarrow j}=-1$.  This is related to the definition in terms of holonomies because for any closed loop $C$, $W(C)$ can be defined as the product of the  $a_{i\to j}$ over all chart overlaps intersecting $C$.}
A theory with global, non-anomalous $P$ symmetry should be well defined in the presence of a non-trivial background for $a$.   A configuration for $a$ may be specified by the co-dimension 1 surfaces across which the orientation changes (see, {\it e.g.},~\cite{Gaiotto:2014kfa}); we refer to these co-dimensional 1 surfaces as parity domain walls. Note that the parity domain walls, as we define them, are not physical objects if parity is an exact symmetry; if parity is spontaneously broken these correspond to physical configurations of localized energy density~\cite{McNamara:2022lrw}.  As we illustrate below, configurations of $a$ with non-trivial parity domain walls lead to color gauge anomalies unless $\theta + \bar \theta = 0$ or $\pi$.

Suppose that we have arbitrary $\theta + \bar \theta$ and are considering the partition function of QCD with a classical background field $a$ that implements orientation reversals across parity domain walls.  We refer to this partition function as $Z_a$.
An example of such a domain wall configuration would be the case where the orientation of space flips as we cross $r =r_\Sigma$ in the spatial radial direction for all time; that is, the parity domain wall $\Sigma$ is a sphere of radius $r_\Sigma$.\footnote{To avoid subtleties about boundaries at $t = \pm \infty$, let us assume that around $t = \pm t_{\rm max}$, for some large time $t_{\rm max}$, the sphere is continuously grown / shrunk from / to a point, so that $\Sigma$ has the topology of $S^3$.}  
We may describe the presence of $\Sigma$ through a unit section $\varepsilon$ of the orientation line bundle, which locally describes the orientation of any point of spacetime. Explicitly, let $\varepsilon=+1$ on ${\mathcal M}_+$ and $-1$ on ${\mathcal M}_-$, with ${\mathcal M}_+$ describing Minkowski space outside $\Sigma$ and ${\mathcal M}_-$ the inside. 
Then, we may write (see also~\cite{Vassilevich:2018aqu} and~\cite{Witten:2016cio} for related discussions)
\es{}{
&\mathrm{exp}\left(i\frac{(\theta+\bar \theta)}{8\pi^2}\int_{{\mathcal M}} \varepsilon \, \mathrm{tr}(G\wedge G)\right)
= \mathrm{exp}\left(i (\theta+\bar \theta) \nu\right) \times \\
&\mathrm{exp}\left(i 2 (\theta+\bar \theta) \frac{1}{8\pi^2}\int_\Sigma \mathrm{CS}_3(A)\right)\,,
}
with 
\es{}{
\mathrm{CS}_3(A) \equiv {\rm tr} \left( A \wedge dA + {2 i g \over 3} A \wedge A \wedge A\right) 
}
the Chern-Simons 3-form localized on $\Sigma$, which is topologically equivalent to an $S^3$, and $\nu$ the usual instanton number, coming from the difference in winding numbers at $t = \pm \infty$.  
Under large color gauge transformations with support on $\Sigma$, $\nu$ is invariant but
\es{}{
\frac{1}{8\pi^2}\int_\Sigma \mathrm{CS}_3(A)\;\to\;
\frac{1}{8\pi^2}\int_\Sigma \mathrm{CS}_3(A)+n \,,
}
for $n \in \mathbb Z$.  To preserve QCD gauge invariance in the partition function $Z_a$,
we must enforce $e^{i 2 n (\theta + \bar \theta)} = 1$, which restricts $\theta + \bar \theta$ to $0$ and $\pi$. Once we enforce $\theta + \bar \theta = 0$ or $\pi$, the partition function $Z_a$ of QCD is well defined for any background field $a$. 

When $\theta + \bar \theta = 0$ or $\pi$ we may gauge parity (option 3 above), which corresponds to summing over all background fields $a$ in the partition function, modulo parity gauge transformations that correspond to continuous deformations of the parity domain walls. After modding out by parity gauge transformations the resulting $P$-gauged partition function is given by~\cite{McNamara:2022lrw,Brennan:2023mmt,Harlow:2023hjb} (see also, {\it e.g.},~\cite{Dijkgraaf:1989pz,Brennan:2023mmt} for similar constructions)
\es{eq:Z_gauged}{
Z_{\rm gauged} \supset \sum_{[a] \in H^1({\mathcal M}, {\mathbb Z}_2)} Z_a \,.
} 
Note that, more precisely, gauging $P$ requires a theory of quantum gravity~\cite{McNamara:2022lrw} (see also App.~\ref{sec:gauge_non_normal_subgroups}), so that the partition function above should be embedded within a full gravitational path integral, which would likely require a sum over all manifolds ${\mathcal M}$, as happens in {\it e.g.} Type-I string theory (see App.~\ref{app:typeI}). However, only the contribution shown in~\eqref{eq:Z_gauged} is relevant for this discussion, and so we suppress the full gravitational path integral throughout this work for simplicity. 
Explicitly, the sum in~\eqref{eq:Z_gauged} is only over manifolds related by non-trivial parity holonomies.  Parity holonomy implies that after being parallel transported along a non-contractable curve a field comes back to itself with a parity transform.  An example of a set of manifolds related by the insertion of a non-contractable parity domain wall are the torus and the Klein bottle. 

Note that the cohomology group $H^1({\mathcal M}, {\mathbb Z}_2)$ is trivial for 4D Minkowski space ${\mathcal M}$; that is, there are no nontrivial parity holonomies because every parity domain wall in Minkowski space can be contracted to a point by parity gauge transformations. 
Thus, the partition function of QCD with gauged parity in 4D Minkowski space looks identical to that of QCD with parity as an exact global symmetry, with both cases requiring $\theta + \bar \theta = 0$ or $\pi$.  On the other hand, option 2---the case of global $P$ symmetry\footnote{See also Ref.~\cite{Vecchi:2025qie}, which defines a symmetry in this way, and also examines which additional assumptions, e.g., on the embedding of the Standard Model in a larger gauge group, would further select $\theta + \bar \theta = 0$.}---is less satisfactory than option 3---the case of local $P$ symmetry---because, as we expand upon below, of the philosophy that all global symmetries, anomalous or non-anomalous, are accidental symmetries in the IR and are not fundamental in the UV.  On the other hand, option 3 provides such a UV motivation for non-anomalous $P$---this is the idea that $P$ is gauged.  In the next section we provide a simple example where gauged $P$ symmetry in 4D Minkowski space descends from a continuous gauge symmetry in a higher dimensional theory with a compact extra dimension. 

Before moving on, we give an additional argument for $\theta = 0$ or $\pi$ for gauged parity, this time in the Hamiltonian formulation. For concreteness, we work in Euclidean signature and compactify time, so that ${\mathcal M} = S^3 \times S^1$; in this case, ${\mathcal M}$ supports non-trivial parity holonomy. We take the interval $[0,1]$ to represent the Euclidean time direction, with $t = 0$ and $t = 1$ identified. However, with gauged parity we must sum over the inclusion of parity domain walls, which here will be inserted at slices of constant $t$:
\es{eq:Z}{
Z = {\rm Tr} e^{-\beta H} + {\rm Tr} P e^{-\beta H} \,,
}
where $H$ is the Hamiltonian, $P$ is the parity operator inserted at $t = 0$, and $\beta = 1$ is the length of the Euclidean time interval. Since the parity operator $P$ can be inserted at any time $t$, it should commute with the Hamiltonian, which sets $\bar \theta = 0$ or $\pi$. On the other hand, we may rewrite~\eqref{eq:Z} as  
\es{eq:Z_rewrite}{
Z = {\rm Tr} (1 + P) e^{-\beta H} \,,
}
which makes it clear that the states need to be even under parity as the odd components are projected out. This restricts the states to the superselection sectors $\theta = 0$ or $\pi$, such that in total $\theta + \bar \theta = 0$ or $\pi$. To make this discussion concrete, in Appendix. \ref{app:1+1QED} we compute the partition function for 1+1D QED with parity gauged, and arrive at the same conclusion that the $\theta$ angle in that theory is restricted to $0$ or $\pi$.

Note that our discussion of gauged parity is completely analogous to how Type I string theory arises from gauging worldsheet parity in Type IIB string theory. In that context, anomaly cancellation leads to the restriction of the gauge group to $\mathrm{SO}(32)$ via the Green–Schwarz mechanism. Here, gauging spacetime parity similarly restricts $\theta$ to $0$ or $\pi$ (see, {\it e.g.},~\cite{Polchinski:1995df}).  The orientifold planes introduced in the case of worldsheet parity are analogous to the parity domain walls discussed above with gauged spacetime parity.  We explain this in more detail in Appendix \ref{app:typeI}.

A UV theory that has gauged $P$ (or gauged $CP$) will then automatically give the `initial condition', from a UV perspective, of $\theta + \bar \theta = 0$ or $\pi$. Of course, $P$/$CP$ must be spontaneously broken at an intermediate energy scale, 
and
it is a model building exercise to ensure that after spontaneous symmetry breaking the neutron EDM is kept small enough for phenomenological purposes while generating the {\it CP} violation in the CKM matrix. On the other hand, this is now an IR model building question that can be addressed at the level of the Lagrangian and not a question about an ambiguity in the superselection sector.  Explicitly, if $P$ or $CP$ is gauged in the UV, then every contribution to the neutron EDM is calculable in terms of the IR physics responsible for spontaneously breaking the discrete symmetry. 

Note  that the spontaneous breaking of $P$/$CP$ can be decoupled from the dynamics involved in gauging the symmetry and take place at much lower energy scales. For example, as we discuss in the next section gauged $P$/$CP$ can arise as the discrete remnant of gauged Lorentz transformations of a higher dimensional gravitational theory compactified on compact extra dimensions. On the other hand, $P$/$CP$ can be spontaneously broken at an energy scale parametrically smaller than the size of those extra dimensions through purely field theoretic mechanisms, though of course $P$/$CP$ could also be spontaneously broken during the compactification. As a simple field theory realization of spontaneously broken $P$ symmetry, let us supplement our theory of pure QCD with an axion field $a(x)$ with the Lagrangian term
\es{}{
{\mathcal L} \supset {-1 \over 32 \pi^2}\left( {a \over f_a} \right) G_{\mu \nu}^a \tilde G^{a \, \mu \nu} \,,
}
with $f_a$ the axion decay constant. Suppose that $a$ has a potential of the form 
\es{}{
V(a) =  -\tfrac12\,\mu^2 a^2 \;+\; \frac{\lambda}{4}\,a^4 \,.
}
This theory is consistent with gauged $P$ with the transformation $a \to - a$ under $P$. On the other hand, $P$ is spontaneously broken by the choice of vacuum $\langle a \rangle = \pm \mu / \sqrt{\lambda}$; in this case, the neutron EDM is calculable and proportional to $\mu / \sqrt{\lambda}$. 

Gauged $P$ or $CP$ provides a clear counterexample to the claims in~\cite{Kaplan:2025bgy}. Moreover, as discussed in {\it e.g.}~\cite{Choi:1992xp}, most discrete solutions to Strong $CP$ discussed previously in the literature can simply be embedded in UV theories that realize $P$/$CP$ as a gauge symmetry, though there are cosmological complications related to stable parity domain walls that should be addressed, for example by limiting the reheat temperature from inflation~\cite{McNamara:2022lrw}.
From a UV perspective gauged $P$/$CP$ is also more motivated than global $P$/$CP$, given
the philosophy that all global symmetries are accidental IR symmetries. In particular, quantum gravity is expected to violate all global symmetries (see, {\it e.g.},~\cite{Banks:2010zn,Harlow:2018tng}). 
If global symmetries are accidental IR symmetries, then there is no easily-justifiable reason to expect $\theta$ to be non-zero. 
Perhaps it is possible to find UV completions where $\theta + \bar \theta = 0$ without tuning, without imposing global symmetries in the UV, and without promoting $P$ or $CP$ to gauge symmetries, but such UV completions are not obvious.   
On the other hand, it certainly is the case that gauged $P$/$CP$ can make this cancellation automatic.

\section{Gauged $P$ from quantum gravity in higher dimensions}   
\label{sec:gauging_CP_QG}

It is well established that in string theory constructions 4D $P$ or $CP$ symmetry arises as a discrete gauge symmetry~\cite{Dine:1992ya,Choi:1992xp,Strominger:1985it,McNamara:2022lrw}.  This is perhaps unsurprising since it is expected that quantum gravity violates all global symmetries~\cite{Banks:1988yz,Harlow:2018jwu,Harlow:2018tng,Abbott:1989jw,Kallosh:1995hi,Chen:2020ojn,Witten:1998qj}. Thus, in embedding $P$ or $CP$   solutions to the Strong $CP$   problem in the context of quantum gravity, it makes sense to suppose that these symmetries are gauged in the UV and spontaneously broken at intermediate energy scales. As we review here, in the context of extra dimensions there are natural embeddings of $P$ and $CP$   into continuous gauge symmetries of the higher dimensional theory. We will begin by presenting an explicit 5D example in which gauged $P$ is realized by embedding $P$ in the proper Lorentz group before reviewing how gauged $P$ and $CP$ arise in specific string compactifications.

\subsection{Gauging $P$: a simple 5D example}
\label{sec:gauged_cp_5D_example}

Here we give a simple 5D example where 4D $P$ arises in the IR as a discrete, unbroken gauge symmetry of a continuous gauge symmetry in the UV through dimensional reduction. Consider a pure 5D $SU(3)$ gauge theory coupled to gravity, with no charged matter, compactified on a circle of radius $R$.  The Lorentz transformation 
\begin{align}
\Lambda_5 = {\rm diag}(1, -1, -1, -1,-1) \,,
\label{eq:Lambda5} 
\end{align}
is continuously connected to the identity.  
We may construct $\Lambda_5$ through a sequence of rotations. First, we perform a rotation by $\phi = \pi$ in the $x_1$-$x_5$ plane, giving the Lorentz transformation matrix $\Lambda = {\rm diag}(1,-1,1,1,-1)$. Next, we perform a rotation by $\pi$ in the $x_2$-$x_3$ plane, giving the rotation $\Lambda_5$.
 Note that we may write $\Lambda_5 = P_4 \times P_5$, with $P_4 = {\rm diag}(1,-1,-1,-1,1)$ the 4D parity operator and $P_5 = (1,1,1,1,-1)$ 5D parity.

Now let us dimensionally reduce the theory on $S^1$. Considering only the Kaluza-Klein (KK) zero modes of the 5D gauge field, this leads to the 4D $SU(3)$ gauge theory with no $\theta$-term in the Lagrangian. 
(As we later discuss, in the presence of the 5D Chern--Simons term, $\theta=\pi$ is allowed.) 
The symmetry $\Lambda_5$ is preserved by the compactification, acting as $x_\mu \to - x_\mu$ and simultaneously $x_5 \to - x_5$, with $\mu = 1,2,3$. Since no fields depend on $x_5$, as we are only considering the KK zero modes, $\Lambda_5$ descends to 4D parity; that is, $P_5$ acts trivially on the states of the EFT. (In theories with fermions this relation is more complicated, as we discuss below and as is discussed in {\it e.g.}~\cite{Dine:1992ya,Choi:1992xp,Strominger:1985it}.)

Let us consider how gauged $P$ arises at the level of the path integral in this scenario using the fact that the Lorentz group of the 5D spacetime is gauged by gravity. In fact, gauging $P$ requires a gravitational theory. A rather elegant way to see this was presented in Ref.~\cite{McNamara:2022lrw}: the $\mathbb{Z}_2$ symmetry group associated to parity is a non-normal subgroup of the full Lorentz group $O(1, d-1)$, so if parity is to be gauged and the (proper) Lorentz symmetry $SO(1, d-1)$ also preserved, the full Lorentz group must be gauged. We justify this in Appendix. \ref{sec:gauge_non_normal_subgroups}
According to general relativity (GR), the metric $g_\mathrm{\mu\nu}(x)$ varies spatially but is locally Minkowski by the equivalence principle, such that there is a local Lorentz symmetry. Of course, for a generic curved spacetime, Lorentz symmetry is not a global symmetry (note also that even for a flat spacetime, global Lorentz symmetry is broken by the compactification in our 5D example).  The local Lorentz symmetry acts in the tangent space of a given point, and is captured by introducing the vielbein fields $e_\mu^a(x)$ such that we may write the metric as
\begin{align}
g_\mathrm{\mu\nu}(x)=e_\mu^a(x) e_\nu^b(x) \eta_{a b} \,,
\end{align}
with $\eta_{a b}$ the Minkowski metric (the indices $\mu,\nu,a,b$ run from $0$ to $d-1$, with $d$ the number of spacetime dimensions).
The local inertial frame is rotated by Lorentz transformations
\begin{align}
e_\mu^a(x) \rightarrow \Lambda^a{ }_b(x) e_\mu^b(x) \,,
\end{align}
where $\Lambda(x) \in \mathrm{SO}(1,d-1)$. This local Lorentz symmetry can be thought of as a gauge symmetry,\footnote{Note that GR is also a gauge theory in a different sense; smooth coordinate-changes $x_\mu \to x'_\mu$ (diffeomorphisms) of the spacetime manifold leave the action invariant and can be viewed as a gauge symmetry \cite{Tong:2009np}; in linearized GR, small diffeomorphisms give a gauge transformation of metric excitation $h$, $h_{\mu \nu} \longrightarrow h_{\mu \nu}+\partial_\mu \xi_\nu+\partial_\nu \xi_\mu$. The graviton, quantum of $h$, is the gauge boson. In this case the gauge group is the infinite-dimensional group of diffeomorphisms, which is not a Lie group as in Yang-Mills. The understanding of GR as a gauge theory has a long and interesting history, see 
\cite{Utiyama:1956sy, 1962rdgr.book..415S,Kibble:1961ba,Bennett:2021dbg,Cartan:1923zea} for details. } with an associated gauge field in the form of the spin connection $\omega^{ab}_\mu(x)$ taking values in in the Lie algebra $\mathfrak{s o}(1, d-1)$. In particular, we can define a covariant derivative on vectors in the tangent space $V^a=e^a_\mu V^\mu$,
\begin{equation}
D_\nu V^a=\partial_\nu V^a+\omega^a{ }_{b \nu} V^b\,,
\end{equation}
provided that the spin connection transforms under $\Lambda$ as 
\begin{align}
\omega \mapsto \Lambda \omega \Lambda^{-1}+\Lambda d \Lambda^{-1}\,.
\label{eq:spin_connection}
\end{align}
From the spin connection we can compute the curvature
\begin{align}
\Omega_{\alpha \beta}{ }^{I J}=2 \partial_{[\alpha} \omega_{\beta]}{ }^{I J}+2 \omega_{[\alpha}{ }^{I K} \omega_{\beta] K}{ }^J .
\end{align}
This is the tetrad formulation of GR \cite{Wald:1984rg,Palatini:1919ffw,Witten:1988hc}. In this formulation the gravitational action reads
\begin{align}
S_\mathrm{tetrad}=\int d^d x\, e \,e_I^\alpha e_J^\beta \Omega_{\alpha \beta}^{I J} \,,
\end{align}
with $e=\sqrt{-g}$, and $g$ is a function of the vielbein. Note that unlike in Yang-Mills theories, the spin connection is not associated to any propagating degrees of freedom because it is not an independent field; it is determined by requiring zero torsion. However, observables must still be gauge invariant in the sense that they must be invariant under local Lorentz transformations.

The path integral of quantum gravity in 5D is of course a difficult object to construct, and at the moment the only known way of making sense of such quantum gravity in 5D and 4D is through the dimensional reduction of string theory starting from a higher dimensional theory.  However, what is clear is that the gravitational path integral on a fixed manifold will include a sum over vielbein fields $e_\mu^a$, up to gauge equivalent identifications. Those gauge equivalent identifications include local Lorentz transformations, including $\Lambda_5 = P_4 P_5$.  In the case in which we only focus on zero modes of the metric and the QCD gauge field, then the gravitational path integral in 5D reduces, in part, to the sum over gauge-inequivalent actions of the operator $P_4$, which gives the partition function written in~\eqref{eq:Z_gauged}.  

From a Hamiltonian perspective,
the sum over Lorentz-equivalent configurations ensures that the partition function is a sum of two pieces that differ by an action of the Lorentz transformation $\Lambda_5 = P_4 P_5$:
\begin{align}
Z &= {\rm Tr} e^{-\beta H} + {\rm Tr} \Lambda_5 e^{-\beta H} \notag \\
&= {\rm Tr}_{S^1} e^{-\beta H_5} {\rm Tr}_{M_4} e^{-\beta H_4} + {\rm Tr}_{S^1} P_5 e^{-\beta H_5} {\rm Tr}_{M_4} P_4 e^{-\beta H_4} \,.
\label{eq:Z_5D_2_4D}
\end{align}
If the fifth dimension is invariant under $P_5$, \eqref{eq:Z_5D_2_4D} reduces to 
\begin{align}
Z = {\rm Tr}_{S^1} e^{-\beta H_5} \left({\rm Tr}_{M_4} e^{-\beta H_4} + {\rm Tr}_{M_4} P_4 e^{-\beta H_4}\right) \,,
\end{align}
where the 4D part is the sum of orientable and non-orientable manifolds.

Let us look at the consequence of $\Lambda_5$ more concretely. In 5D, there is no $\theta$-term but there can be the Chern--Simons (CS) term in addition to the usual gauge kinetic term,
\begin{align}
    {\cal L}_{\rm gauge}
    &= - \frac{1}{2g_5^2} {\rm tr} G_{\mu\nu} G^{\mu\nu}
    + \frac{k}{24\pi^2} \epsilon^{\mu\nu\rho\sigma\kappa} {\rm tr} 
    \left(\frac{1}{4} A_\mu G_{\nu\rho} G_{\sigma\kappa} 
    \right. \nonumber \\
    &\phantom{=} \left.
    - \frac{1}{4} A_\mu A_\rho A_\sigma G_{\sigma\kappa} 
    + \frac{1}{10} A_\mu A_\nu A_\rho A_\sigma A_\kappa\right)
    \nonumber \\
    &\phantom{=} + \mbox{dimension six operators and above} \,,
    \label{eq:CS}
\end{align}
where here $A_\mu$ the $SU(3)$ gauge field and $G_{\mu\nu}$ its field strength. The CS term is the only one in the pure Yang--Mills theory that can potentially lead to the $\theta$ term in 4D.
The coefficient $k \in {\mathbb Z}$ above must be an integer to maintain the gauge invariance as the CS term changes by an integer under large gauge transformations. The CS term breaks the 5D parity explicitly, but of course is 5D Lorentz invariant and hence preserves $\Lambda_5$. Compactifying on $M_4 \times S^1$, where $S^1$ has the circumference $R$, the first term in \eqref{eq:CS} gives the usual 4D Lagrangian with $g_4^2 = g_5^2 R^{-1}$. If the gauge group breaks by a Wilson line $W=e^{i \oint_{S^1} A_5^a T^a d x^5} \neq I$ for one of the generators $a$, the CS term can induce the $\theta$-term in 4D and hence violate the 4D parity $P_4$. However, if we assume the compactification preserves $P_5$, under which $W \rightarrow W^{-1}$, then we need every component of $W = W^{-1} = (-1)^n$ and hence $A_5 = n \frac{\pi}{R}$ for $n \in {\mathbb Z}$. The only term that survives in \eqref{eq:CS} is the first one in the parentheses where $A_5$ may be part of $A_\mu$ or $G_{\nu\rho}$ or $G_{\sigma\kappa}$ with three possible choices. Integrating over $S^1$ yields
\begin{align}
\lefteqn{
    \oint_{S^1} \frac{k}{24\pi^2} \epsilon^{\mu\nu\rho\sigma\kappa} {\rm tr} 
    \left(\frac{1}{4} A_\mu G_{\nu\rho} G_{\sigma\kappa} - \frac{1}{4} A_\mu A_\nu A_\rho G_{\sigma\kappa} 
    \right. } \nonumber \\
    & \left.
    + \frac{1}{10} A_\mu A_\nu A_\rho A_\sigma A_\kappa \right)
    = \frac{kn\pi}{8\pi^2} \epsilon^{\nu\rho\sigma\kappa} {\rm tr} 
    \frac{1}{4} G_{\nu\rho} G_{\sigma\kappa}  \,,
\end{align}
which implies $\theta = k n \pi$, preserving $P_4$. $A_5$ will acquire mass from higher order corrections and it is quite possible that the minima are at $A_5=0, \pi/R$ given that they are symmetry-enhanced points. 

When considering fermions
\begin{align}
    {\cal L}_{\rm fermion}
    &= \bar{\psi}(i \gamma^\mu D_\mu - m) \psi,
\end{align}
the mass $m$ breaks the 5D parity and must be real to ensure the hermiticity of the Lagrangian. It is impossible to define charge conjugation symmetry in 5D. The $\Lambda_5$ acts on the Dirac spinor $\psi(x^0, x^{1,2,3,5}) \rightarrow i \gamma^1 \gamma^2 \gamma^3 \gamma_5 \psi(x^0, -x^{1,2,3,5})$ where we use the 4D notation with anti-hermitian $\gamma^{1,2,3}$ and hermitian $\gamma_5$. The product is nothing but $\gamma^0$ which defines the usual $P_4$. Note that the fermion Lagrangian could have contained a complex mass term $-m \bar{\psi}\psi  \cos\phi - i m \bar{\psi} \gamma_5 \psi \sin\phi$ if only 4D Lorentz invariance is imposed, while the invariance under $\Lambda_5$ forbids the second term. If there were the second term, a chiral rotation would eliminate it and also induce the $\theta$-term. Once again the invariance under $\Lambda_5$ achieves the 4D parity symmetry. Note, however, that the resulting $\theta$ may be $0$ or $\pi$ depending on the sign of the 5D mass.

This construction can be generalized to higher odd dimensions, for instance on $M_4 \times S^1 \times S^{2n}$, if the 4D Dirac fermion is obtained due to the flat connection on $S^1$ and the index $c_n \equiv \frac{1}{(2 \pi)^n n!} \operatorname{tr} F^n$ on $S^{2n}$. When $\oint_{S^1} A = \pi$, we find the CS term $\omega_{2n+5}$, defined via $\mathrm{d}\omega_{2n+5}=c_{n+3}$, to be $\omega_{2n+5} = \frac{1}{2} c_n c_3$, and again the 4D $\theta$ angle is $0$ or $\pi$. The 4D mass term is also real.
In higher $4+2n$ dimensions, we can have a $c_{2+n}$ term with an arbitrary coefficient in the Lagrangian. In order to obtain fermion zero modes, we need an index $c_n$, and hence a 4D $\theta$ term. 

\subsection{Gauging $P$ and $CP$ in string theory}
\label{sec:gauging_P_ST}

We have seen that gauging $P$ requires a theory of quantum gravity. 
In string theory, if $P$ or $CP$ are symmetries, they are expected to be gauged. In the following we review how $P$ and $CP$  can survive as 4D gauge symmetries after compactification in string theory. (See also Refs. \cite{Feruglio:2023uof,Baur:2020yjl}, who recently suggested a string-inspired solution to the Strong $CP$ problem using modular invariance.)

As we see in the example of the previous subsection, $P$ is gauged if it may be embedded as an unbroken element of a higher dimensional Lorentz group. We have provided a simple construction to perform this embedding for odd-dimensional spacetimes.
Generalizing slightly the discussion in \ref{sec:gauged_cp_5D_example}, the trick was to note that in $2n$-dimensional Minkowski space 
\begin{equation}
\left(x^0, x^1, \ldots, x^{2 n-1}\right) \longrightarrow\left(x^0,-x^1, \ldots,-x^{2 n-1}\right)
\end{equation}
is in the component of the Lorentz group which is disconnected from the identity, which is in the same disconnected component as the transformation $x^1 \to -x^1$.
If an extra spatial dimension $x^{2n}$ is introduced, this $Z_2$ symmetry is embedded in continuous $SO(2)$ rotation group of $(x^{2n-1}, x^{2n})$, so it is a proper Lorentz transformation in $2n+1$ dimensions.  The same trick works in some string constructions. For example, consider a toroidal compactification in 10D string theory on $M_4 \times K$ with $M_4$ 4D Minkowski space and the internal manifold given by $K=T^2 \times T^2 \times T^2$, with $T$ the torus. Grouping the internal coordinates as $z^j=x^j+ix^{j+1}$ for $j=4,6,8$, the transformation $\Lambda$ which combines 4D parity with the product of the conjugations $z^j \to \bar{z}^j$ is a proper Lorentz transformation of the 10D theory. Note that it reverses the orientation of $K$ because an odd number of internal coordinates flip sign, and also of $M_4$, but preserves the orientation of the full manifold. $\Lambda$ also commutes with the Gliozzi–Scherk–Olive condition \cite{Gliozzi:1976qd}, which ensures consistency with the wordlsheet supersymmetry and the physical spectrum. By examining the 10D Dirac algebra $\Lambda$ can be shown to flip the handedness of massless fermions, so it acts as 4D parity $P$ \cite{Dine:1992ya}. This embedding of $P$ is interesting because it applies to, {\it e.g.}, 10D heterotic string theory, which is not invariant under 10D parity to begin with (the theory contains Majorana-Weyl spinors of a definite handedness).  For a compactification on a more complicated manifold such as a Calabi-Yau (CY) 3-fold, the same embedding of parity above can work: it would suffice to find an isometry $\mathcal{I}_K$ of $K$ (\textit{i.e} a diffeomorphism of $K$ which preserves the metric) which reverses the orientation of $K$. When combined with 4D parity, the orientation of the 10D manifold is preserved. For a CY 3-fold any such  $\mathcal{I}_K$ must be anti-holomorphic, such as $z^j \to \bar{z}^j$. In general $z^j \to \bar{z}^j$ is not a isometry of the internal space, though for certain CY it can be \cite{Strominger:1985it}. For example, the quintic hypersurface in $\mathbb{C}P^4$
\begin{equation}
\sum_{i=1} z_i^5+\varepsilon z_1 z_2 z_3 z_4 z_5=0
\end{equation}
is a CY 3-fold. Then $z_i \to \bar{z_i}$ is a symmetry  if and only if $\varepsilon$ is real.  Lastly, note that in some string theory examples, such as 11D M-theory  \cite{Witten:2016cio}, the higher-dimensional theory is already parity-symmetric. 

Let us now ask how $CP$ can arise as a gauged symmetry in string theory.  As discussed in Ref.~\cite{Choi:1992xp} (see also \cite{McInnes:1995hj} for a more mathematical perspective), suppose we wish to gauge $CP$  by realizing it as a product of transformations $CP=X_L X_\mathrm{YM}$ with $X_L$ an element of the $d$-dimensional Lorentz group, and $X_{Y M}$ an element of the internal Yang-Mills group $G_\mathrm{YM}$ (which must be nontrivial to have chiral fermions). Note that there may be other ways to gauge $CP$. Supposing that all 4D fields live in an irreducible representation of $G_\mathrm{YM} \times SO(d-1,1)$, these representations must be self-conjugate. This is possible if $X_L$ ($X_\mathrm{YM}$) is an inner automorphism \footnote{\textit{i.e} is a map from the group to itself $x \to gxg^{-1}$ for some group element $g$}
of $SO(d-1,1)$ ($G_\mathrm{YM}$) which exchanges an element of a representation with its complex conjugate. 
This requires irreducible spinor representations to be real or pseudo-real.
As spinor representations are complex in 4D, at least five spacetime dimensions are needed to gauge $CP$ in this way. Restricting to real representations requires $8k+1$, $8k+2$, or $8k+3$ spacetime dimensions (for $k$ integer).
If one allows for Majorana spinors, which exist in $8k+2$ dimensions (further, in these dimensions Majorana-Weyl spinors exist), the minimal dimension to gauge $CP$ as above becomes 10 \footnote{We believe the discussion in Ref.~\cite{Choi:1992xp} contains errors.  First, there it is claimed that gauging $CP$ by realizing it as a product $CP=X_L X_\mathrm{YM}$ requires 4D fields to be in irreducible representations of $G_\mathrm{YM} \times SO(d-1,1)$. However, a priori there is no reason to not consider reducible representations instead. Ref.~\cite{Choi:1992xp} also implicitly assumes that self-conjugate irreducible spinor representations must be Majorana, and further that Majorana spinors exist in $8k+1$ dimensions. This is incorrect: a self-conjugate representation is either real or pseudo-real; in Minkowski space Majorana spinors exist only in $8k+2$, $8k+3$, and $8k+4$ spacetime dimensions  \cite{Slansky:1981yr}.  }. 
In general $G_\mathrm{YM}$ does not admit inner automorphisms which flip the sign of the entire Cartan subalgebra \footnote{A group element $g \in G_\mathrm{YM}$ acts on the Lie algebra via the adjoint action $\mathrm{Ad}_g(X)=gXg^{-1}$. We are interested in a $g$ for which $\mathrm{Ad}_g(X)$ transforms the generators of the Cartan subalgebra as $H_i \to - H_i$, and the step operators as $E_\alpha \to E_{-\alpha}$. This ensures that the action of $g$ flips the signs of weights $\lambda \to -\lambda$, meaning that the action of $g$ realizes an isomorphism between any representation of $G_\mathrm{YM}$ to its conjugate representation.}; the only Lie groups which do are $E_8$, $E_7$, $SO(2n+1)$, $SO(4n)$, $Sp(2n)$, $G_2$, $F_4$, and products of these groups do \cite{Slansky:1981yr,Henning:2021ctv}. Remarkably, this means that heterotic or Type I string theory with gauge group $E_8 \times E_8$ or $SO(32)$ can realize $C$ as a gauge symmetry of the 10D theory. In particular, compactifying on a CY 3-fold $K$, an orientation-reversing isometry $\mathcal{I}_K$ of $K$ can sometimes be identified as $C$. For example, in $E_8 \times E_8$ heterotic string theory, the spin connection on the tangent bundle of $K$, which has complex structure group $SU(3)$, can be embedded in the gauge connection of a single $E_8$ via $SU(3) \hookrightarrow S U(3) \times E_6 \hookrightarrow E_8$. The result is that in 4D the  $E_8$ gauge group is broken to $E_6$, and $\mathcal{I}_K$ exchanges the $\mathbf{27}$ and $\mathbf{\overline{27}}$ representations of $E_6$ \cite{Candelas:1985en}, so it acts as $C$ \footnote{This deserves some explanation. Under $E_8 \supset E_6 \times SU(3)$, the adjoint decomposes as $\mathbf{248}\to(\mathbf{78},\mathbf{1})\oplus(\mathbf{1},\mathbf{8})\oplus(\mathbf{27},\mathbf{3})\oplus(\overline{\mathbf{27}},\overline{\mathbf{3}})$. Since $K$ has $SU(3)$ holonomy, the spin connection on the tangent bundle of $K$ (see \eqref{eq:spin_connection}) is reduced to values in $
\mathfrak{s u}(3) \subset \mathfrak{s o}(6) $. It can thus be identified as the gauge connection of the $SU(3)$ subgroup of $E_8$. In this case, an orientation-reversing isometry, being anti-holomorphic, exchanges the $\mathbf{3} \leftrightarrow \bar{\mathbf{3}}$ and thus also the gauge group representations $\mathbf{27} \leftrightarrow \overline{\mathbf{27}}$.}. In this case combining $\mathcal{I}_K$ with 4D parity realizes $CP$ as a gauge symmetry in 4D, with $CP$ of the form $X_LX_\mathrm{YM}$ as discussed above. 

A common place where gauged parity is discussed is in the context of Type I string theory, which implements gauged worldsheet parity. We discuss this case in detail in App.~\ref{app:typeI} and draw parallels between that well-understood case and our discussion above about gauged parity in larger spacetime dimensions.

\section{There is a Strong $CP$   problem}
\label{sec:is_strong_cp}

In this Section we respond to recent claims in the literature that there is no Strong $CP$   problem, \textit{i.e} that the neutron EDM vanishes even if $\bar{\theta }\neq 0$. We focus on the claim of Ref.~\cite{Ai:2024cnp} that boundary conditions in the path integral set the neutron EDM to zero.

\subsection{Order of limits in QCD partition function}
\label{app:there_is_strong_cp_prob}
In Ref. \cite{Ai:2024cnp} (see also \cite{Ai:2020ptm,Ai:2022htq}) it is claimed that $CP$ is conserved in QCD, and therefore the neutron EDM vanishes, due to a subtlety in the order of limits one takes when computing the expectation value of local operators. 
In \eqref{eq:PathIntegral}, we sum over topological sectors labeled by $\nu$ and send $\nu \to \infty$; implicitly we also take the limit of infinite volume, $V \to \infty$. The claim of Ref.~\cite{Ai:2024cnp} is that the only mathematically sound procedure is to first take $V \to \infty$ when computing the action in a given sector $\nu$ and then $\nu \to \infty$. This would lead to a vanishing neutron EDM, as well as a vanishing topological susceptibility of QCD, independently of $\bar{\theta}$ \footnote{In later work by the same authors Ref.~\cite{Ai:2024vfa}, a similar argument is presented that does not rely on taking the limit of infinite-volume. There it is claimed that the theta-parameter of pure Yang--Mills theory on the 4D Euclidean torus does not have physical effects. This claim is incorrect for the same reasons we discuss. Note that in Appendix \ref{app:1+1QED} we present a finite-volume construction where the theta-term clearly has physical effects, 1+1D QED on a circle. }. Recall that the topological susceptibility of QCD is given by
\begin{align}
&\left.\chi_{\text {QCD }} \equiv \frac{\partial^2 E(\theta)}{\partial \theta^2}\right|_{\theta=0}=\int d^4 x\langle  q(x) q(0)\rangle\,,
\end{align}
with $E(\theta)$ the vacuum energy and $q(x)=\frac{g^2}{32 \pi^2} F \tilde{F}$ the topological charge density.

Ref.~\cite{Albandea:2024fui} pointed out that this claim can be understood to be incorrect by analogy to a simpler setting, the quantum mechanical example of the particle on a ring (see also App.~\ref{app:1+1QED}). They consider a particle of mass $m$ moving on a circle of unit radius, with Lagrangian
\begin{align}
L = \frac{1}{2} m\dot \phi^2 + \frac{\theta}{2 \pi} \dot \phi \,,
\label{eq:particleonring_main_text}
\end{align}
where $\phi$ is the coordinate of the particle on the ring and there is a total derivative in analogy with the $\theta$ term of QCD. In this system time plays the role of the volume for QCD and one could work in Euclidean time $\tau=it$ with periodic boundary conditions $\tau \sim \tau +\beta$. We can define a topological charge
\begin{align}
Q=\frac{1}{2 \pi} \int_0^\beta 
d \tau \frac{d{\phi}}{d\tau}
\in \mathbb{Z}\,
\label{eq:Qdef}.
\end{align}
The partition function and energy levels of the system are
\begin{align}
Z(\theta)=\sum_{n \in \mathbb{Z}} e^{-\beta E_n}\,,\quad E_n = \frac{1}{2 m}\left(n-\frac{\theta}{2 \pi}\right)^2.
\end{align}
The expectation value of an operator $\mathcal{O}$ is given by
\begin{align}
\langle\mathcal{O}\rangle=\lim _{N \rightarrow \infty} \lim _{\beta \rightarrow \infty} \sum_{|Q|<N}\langle\mathcal{O}\rangle_Q p(Q) \,,
\end{align}
where $\langle\mathcal{O}\rangle_Q$ indicates the expectation value fixing the topological sector $Q$, and 
where the topological charge density is 
\begin{align}
p(Q) & =\frac{1}{2 \pi Z(0)} \int_{-\pi}^\pi d \theta Z(\theta) e^{-i \theta Q} 
\\
&=\frac{1}{Z(0)} \sqrt{\frac{2 m \pi}{\beta}} \exp \left(-\frac{2 m \pi^2}{\beta} Q^2\right) \,.
\end{align}
If we compute the topological susceptibility in this theory,
\begin{align}
\chi_t= \langle Q^2\rangle = \lim _{N \rightarrow \infty} \lim _{\beta \rightarrow \infty} \frac{1}{\beta} \frac{\sum_{|Q|<N} Q^2 p(Q)}{\sum_{|Q|<N} p(Q)} \,,
\label{eq:sum_topo}
\end{align}
taking first the limit $\beta \to \infty$, we obtain $\chi_t=0$. On the other hand there is no issue with computing the topological charge at finite $\beta$; the sum over $Q$ in \eqref{eq:sum_topo} is perfectly well-defined for any value of $\beta$, and taking first the limit $N\to \infty$ we get $\chi_t=1/(4\pi^2\, m)$, which is the correct result that may be computed directly from the two-point function of $Q$, using~\eqref{eq:Qdef}, without any ambiguous order of limits~\cite{Albandea:2024fui}.

As in the example above, 
the order of limits claimed to be correct in Ref.~\cite{Ai:2024cnp} would set the topological susceptibility to zero in pure Yang-Mills, just as it would in QCD (this is claimed to be the case in~\cite{Ai:2024cnp}), which is clearly incorrect. In the limit of large $N_c$ (number of colors), it can be proved that at leading order, the mass squared of the $\eta'$ meson is proportional to the pure Yang-Mills topological susceptibility through the Witten-Veneziano relation \cite{Witten:1979vv,Veneziano:1979ec}
\begin{align}
m_{\eta^{\prime}}^2\approx
m_\eta^2+m_{\eta^{\prime}}^2-2 m_K^2=\frac{4 N_f}{f_\pi^2} \chi_\mathrm{YM} \,,
\label{eq:WV_main}
\end{align} 
with $m_K$ and $m_{\eta}$ respectively the mass of the Kaon and $\eta$ meson. The mass of the $\eta'$ is measured to be $m_\mathrm{\eta'}\, \approx 957\, \mathrm{MeV}$, and thus $\chi_\mathrm{YM}\neq 0$ \footnote{A argument using current algebra for why a massive $\eta'$ directly implies violation of $CP$ in QCD is given in  \cite{Shifman:1979if}.}. The relation in \eqref{eq:WV_main} has been confirmed by lattice calculations \cite{DelDebbio:2004ns,Cichy:2015jra}. While this holds in the large $N_c$ limit, it reproduces remarkably well the measured value of  $m_{\eta'}$, using the value of $\chi_\mathrm{YM}$ extracted from pure $SU(3)$ lattice computations, $\chi_\mathrm{YM}\approx 190$ MeV \cite{DelDebbio:2004ns}. Note that \eqref{eq:WV_main} can be derived without needing to discuss the order of the limits in the partition function (and also independently of the chiral Lagrangian), as we summarize in Appendix.~\ref{app:WV_derivation}.

\subsection{$\bar\theta$ in the Chiral Lagrangian}
\label{app:error_Tamarit_chiral_PT}

That $\bar{\theta}$ has physical effects, \textit{e.g.}, on the neutron EDM or the vacuum energy of QCD, is well established in $\chi$-PT \cite{Srednicki:2007qs,Baluni:1978rf,Crewther:1979pi}. The claim of Ref. \cite{Ai:2024cnp} would imply that this textbook result is incorrect, despite $\chi-\mathrm{PT}$, being an effective theory of pseudoscalar mesons, knowing nothing of the topological sectors of QCD other than through Wilson coefficients. Below, we summarize the standard derivation of this result, and then point out the inconsistency in  Ref.~\cite{Ai:2024cnp}, which argues that the chiral Lagrangian is independent of $\bar{\theta}$.

Recall that in the limit of massless quarks, QCD with $N_f$ flavors possesses the flavor symmetries $S U\left(N_f\right)_L \times S U\left(N_f\right)_R \times U(1)_V$. At low energies, the chiral symmetry is broken down to the diagonal subgroup $S U\left(N_f\right)_L \times S U\left(N_f\right)_R \rightarrow S U\left(N_f\right)_{\text {diag }}$ by the vacuum expectation value (VEV) of the quark condensate $\langle\bar{q} q\rangle \neq 0$. The leading order terms of the chiral Lagrangian consistent with these symmetries are

$$
\mathcal{L}_{\mathrm{M}}^{\mathrm{EFT}}=\frac{f_\pi^2}{4} \operatorname{Tr} \partial_\mu U \partial^\mu U^{\dagger}+\frac{f_\pi^2 B_0}{2} \operatorname{Tr}\left(M U+U^{\dagger} M^{\dagger}\right),
$$
with $U \in S U\left(N_f\right)$ the matrix of pseudo-Goldstone bosons of the chiral symmetry breaking, $f_\pi$ the pion decay constant, and $M=\operatorname{diag}\left(m_q\right)$ the mass matrix, with $m_q$ the quark masses, and $B_0$ a constant. Note that we may parametrize $U=\exp \left(i \pi^a \tau^a\right)$ with $\pi^a$ the $N_f^2-1$ Goldstone fields and $\tau^a$ the generators of $S U\left(N_f\right)$ normalized such that $\operatorname{Tr}\left[\tau^a \tau^b\right]=2 \delta^{a b}$. Consider the topological term in the QCD Lagrangian $ \mathcal{L} \supset-\frac{\bar{\theta}}{32 \pi^2} G \tilde{G}$.
The chiral anomaly ensures that this term is removed after performing a chiral rotation of the quark fields $q \rightarrow e^{i \gamma_5 \bar{\theta} / 4} q$, corresponding to a rotation of the mass matrix $M \rightarrow e^{i \bar{\theta} / 2} M$. Restricting to the two lightest quarks $u, d$ for simplicity, a quick computation shows that $U$ acquires a VEV of the form $\langle U\rangle=\operatorname{diag}\left(e^{i \phi}, e^{-i \phi}\right)$ with
$$
\tan \phi=\frac{m_d-m_u}{m_d+m_u} \tan \frac{\bar{\theta}}{2} \,.
$$
The vacuum energy then depends on $\bar{\theta}$ as
$$
E(\bar{\theta})=-f_\pi^2 m_\pi^2 \sqrt{1-4 \frac{m_u m_d}{\left(m_u+m_d\right)^2} \sin ^2 \frac{\bar{\theta}}{2}} \,.
$$
The chiral Lagrangian contains terms of the form
\begin{align}
\mathcal{L} \supset-c_3 \operatorname{tr}\left[M U+M^{\dagger} U^{\dagger}\right] \bar{N}\left(U^{\dagger} P_L+U P_R\right) N \,, 
\end{align}
with $N=(p, n)^T$ the proton-neutron doublet under $\mathrm{SU}(2)$ isospin and $P_L\left(P_R\right)$ left (right) handed projectors. This term clearly depends on $\bar{\theta}$ after the rotation of $M$; expanding at small $\bar{\theta}$ we obtain a term $\mathcal{L} \sim \bar{\theta} \pi^{+} \bar{p} n$ which at one loop gives the neutron EDM $d_n \approx 3 \times 10^{-16} \bar{\theta}\, e \cdot \mathrm{~cm}$.

To argue instead that this $\bar{\theta}$-dependence is not present, Ref.~\cite{Ai:2024cnp} follows an alternate, heuristic approach to extend the chiral Lagrangian to include the $\eta'$  (see \textit{e.g}, \cite{Witten:1978bc,Witten:1980sp, Veneziano:1979ec,Csaki:2023yas}), which we now discuss. 
Let us consider $N_f=3$. At the \textit{classical} level, the massless QCD Lagrangian has a $U(3)_L\times U(3)_R$ chiral symmetry and there are nine Goldstone bosons associated with the spontaneous symmetry breaking to the diagonal subgroup $U(3)_V$. These Goldstone bosons can be parametrized by the $3\times 3$ unitary matrix
\begin{equation}
 U = U_0 \exp\left[\frac{i}{f_\pi}\left( \frac{\eta_1}{\sqrt{3}}I_3 + \frac{\lambda_i}{\sqrt{2}}\phi_i\right)\right]\,,
\end{equation}
where $\lambda_i$ are the Gell-Mann matrices and the $U(1)\times SU(3)$ content of the broken $U(3)_A$ is explicit. Under the chiral group, $U$ transforms as $g_R\, U\, g_L^{\dagger}$, with $g_{R,L} \in U(3)_{R,L}$. The quark mass matrix $M$ explicitly breaks $U(3)_A$.

However, the $U(1)_A$ part of the chiral symmetry group is anomalous in QCD. A naive approach to account for the chiral symmetry breaking effect induced by the anomaly in the eﬀective low-energy theory, would be through the term
\begin{equation}
|\lambda| \mathrm{e}^{-\mathrm{i} \xi} f_\pi^4 \operatorname{det} U + \mathrm{h.c} \,.
\label{eq:U1A_effective_term}
\end{equation}
This is the route followed by the authors of Ref.~\cite{Ai:2024cnp}. Treating $\theta$ and $M$ as spurions, they should transform properly under $U(1)_A$, as
\begin{align}
\begin{split}
&M \rightarrow \mathrm{e}^{-2 \mathrm{i} \beta} M,\,  \theta \rightarrow \theta+2 N_f \beta \,,\\
&\operatorname{det} U \rightarrow \mathrm{e}^{2 \mathrm{i} N_f \beta} \operatorname{det} U \,.
\label{eq:U1A}
\end{split}
\end{align}
Ref.~\cite{Ai:2024cnp} claims that setting $\xi = - \mathrm{arg}\,\mathrm{det}M $ is consistent with these transformations and is in fact the only option for $\xi$ that respects their conclusion about the order of limits in the partition function, leaving the chiral Lagrangian independent of $\theta$. In fact, a chiral rotation on $U$ that would make $M$ real, would remove every phase from the Lagrangian. 
However, this choice for $\xi$ is not consistent with QCD as it implies that the EFT is invariant under the simultaneous rotations of $M$ and $U$ alone  (given in \eqref{eq:U1A}) without any shift in $\theta$, whereas this is not a symmetry of QCD.

Moreover, that this conclusion is incorrect can be seen via a more consistent (but still heuristic) construction of the low energy Lagrangian from QCD. In fact, requiring the correct transformation under a $U(1)$ axial rotation, the absence of color degrees of freedom (confinement) and demanding that the effective Lagrangian follows from QCD in the large-$N_c$ limit, at leading order in $1/N_c$, one gets~\cite{DiVecchia:1980yfw}
\begin{align}
\begin{split}
\mathcal{L}_{\rm EFT} \supset &\frac{f_\pi^2}{4} \operatorname{Tr} \partial_\mu U \partial^\mu U^{\dagger}+\frac{f_\pi^2 B_0}{2} \operatorname{Tr}\left(M U+U^{\dagger} M^{\dagger}\right)\\
&\frac{i}{2}q(x)\Tr[\log U - \log U^{\dagger}] + \frac{N_c}{a f_{\pi}^2}q^2(x) -\bar{\theta} q(x)\,,
\end{split}
\end{align}
where $q(x) = \frac{g^2}{64\pi^2}G^{\mu\nu}\tilde{G}_{\mu\nu}$, $a$ is a constant that is of order one as $N_c\to \infty$, the normalization is chosen for convenience and we work in the basis where $M$ is real, so that all phases are encoded in $\bar{\theta}$. Removing $q(x)$ with its equation of motion, we get
\begin{align}
\begin{split}
\mathcal{L}_{\rm EFT} \supset &\frac{f_\pi^2}{4} \operatorname{Tr} \partial_\mu U \partial^\mu U^{\dagger}+\frac{f_\pi^2 B_0}{2} \operatorname{Tr}\left(M U+U^{\dagger} M^{\dagger}\right)\\
&-\frac{a\, f_{\pi}^2}{4N_c}\bigg(\bar\theta - \frac{i}{2}\Tr[\log U - \log U^{\dagger}]\bigg)^2\,.
\label{eq:correct_chiral_EFT}
\end{split}
\end{align}
Consequently after expanding, we see that $m_{\eta'}^2$ scales as $1/N_c$. Note that \eqref{eq:correct_chiral_EFT} can brought in a form similar to \eqref{eq:U1A_effective_term} since $\Tr(\log U) = \log\operatorname{det}U$. The logarithm in \eqref{eq:correct_chiral_EFT} is needed as we know from $1/N_c$ expansion counting rules that the anomaly-induced interaction is, to leading order in 1/$N_c$, quadratic in $\eta_1$ only~\cite{Witten:1980sp}. It is therefore evident that the low energy theory which encodes all features of QCD has a physical dependence on $\bar{\theta}$. A chiral rotation on $U$ would only shift it between $M$ and the last term in \eqref{eq:correct_chiral_EFT}.  \\

\section{Discussion}
\label{sec:discussion}

In this work we show explicitly that (i) the Strong $CP$ problem is a real problem and not trivially solved by {\it e.g.} the neutron EDM having no dependence on $\bar \theta$ (it does), and (ii) that models based on gauged but spontaneously broken $P$ or $CP$ can solve this problem. Note that our statement (ii) is an ``in principle'' statement in that we do not develop new models solving the Strong $CP$ problem through discrete symmetries but rather we show that many existing solutions can be embedded in UV models of gauged $P$/$CP$.  While this has largely been known for decades (see, {\it e.g.},~\cite{Choi:1992xp}), it bears repeating in light of the recent work~\cite{Kaplan:2025bgy}. Contrary to the claims in~\cite{Kaplan:2025bgy} (who claim that only the axion can solve the Strong $CP$ problem), models in which $P$ or $CP$ is gauged in the UV do not suffer from any ambiguity in theta state versus Lagrangian contributions to the neutron EDM; in these cases, $\theta + \bar \theta = 0$ or $\pi$ in the UV, with $\theta$ representing the contribution to the neutron EDM from the theta vacuum and $\bar \theta$ denoting the Lagrangian contribution. After spontaneous $P$ or $CP$ symmetry breaking, the contributions to the neutron EDM are calculable from the physics responsible for the breaking. (This, of course, does not mean that constructing such models based off of gauged $P$ or $CP$ in the UV that satisfy all phenomenological requirements is easy.)

On the other hand, at a deeper level it is unsatisfying to imagine a theory where the neutron EDM depends on a parameter $\theta$ describing the theta state that is otherwise incalculable by the UV completion of the theory.  We conjecture that in theories that arise from quantum gravitational theories in the UV, the IR value of $\theta + \bar \theta$, which is the only physically observable combination of theta angles (including, of course, the contribution from the quark mass phases in $\bar \theta$), is dynamically determined from the UV theory. Our reasoning for this conjecture is the following. 

The key point is that the existence of theta vacua in a non-abelian gauge theory is tied to the existence of a global symmetry; in particular, a Chern-Weil global symmetry~\cite{Heidenreich:2020pkc}.  The operator $\tr (F \wedge F)$ is a closed form and thus generates a generalized global symmetry; in the case of 4D Yang-Mills theory this is a $(-1)$-form global symmetry. Since quantum gravity is conjectured to have no exact global symmetries, it follows that this $(-1)$-form global symmetry should be gauged or broken. Gauging the Chern-Weil global symmetry corresponds to introducing an axion field $a$, which has an exact shift symmetry (which is the gauge symmetry). The axion is the gauge boson of the $(-1)$-form gauge symmetry~\cite{Heidenreich:2020pkc}. Since the axion shift symmetry is a gauge redundancy, the concept of the theta vacua is also redundant; we can shift the axion field to cancel any base contribution $\theta + \bar \theta$. Thus, 
a unique and calculable value of $\theta + \bar \theta$ 
is generated in the IR.\footnote{There is however an alternate possibility, emphasized in \cite{Reece:2023czb,Heidenreich:2020pkc,Cecotti:2018ufg}, which is to break the Chern-Weil global symmetry. This corresponds to the case where $\theta$ is ``frozen" in the IR to one of a discrete set of possible values, without any modulus (such as an axion) that can vary it away from those values. 
For example, in Type IIB string theory compactified on a rigid Calabi-Yau 3-fold, there is no light axion. Even in this example, the physical theta parameter in the IR is calculable, and in fact must be $0$ or $\pi$. Ref.~\cite{Reece:2023czb} however argues that all known examples which break the Chern-Weil symmetry lack light charged matter and thus cannot be phenomenologically viable.}

Of course, just because the theory has an axion does not mean that the axion solves the Strong $CP$ problem. For example, the axion could spontaneously acquire a potential through {\it e.g.} Euclidean $D$-branes wrapping extra dimensions in a compactification of a higher dimensional theory or through the confinement of a hidden gauge sector with a confinement scale well above that of QCD. This would make the axion heavy and, generically, introduce a non-zero $\theta + \bar \theta$ in the IR EFT.\footnote{It is plausible that the Strong $CP$ problem is solved by a symmetry-based solution even if a heavy QCD axion is present. For example,  corrections to the axion potential may preserve $P$/$CP$ due to the geometry of the compactification, see \textit{e.g,} \cite{Ishiguro:2020nuf, Bonisch:2022slo,Kobayashi:2020uaj}.}  However, in this case the value of $\theta + \bar \theta$ in the IR is fully calculable from the physics that sets the axion's potential and does not depend on an unknowable superselection sector. Thus, in principle, with creative model building the phase of this potential could be aligned with that of QCD, such that $\langle a / f_a \rangle \sim 0$. (To emphasize the point, though, just because this is in principle possible does not mean that constructing such a model is easy.)

Indeed, in the one theory of quantum gravity that we have---string theory---axions are abundant in 4D EFTs, where they often arise from the dimensional reduction of higher-form gauge fields during compactification to 4D. Those higher-dimensional higher-form gauge fields can often be thought of as gauging the Chern-Weil global symmetries that would otherwise be present. The higher-form gauge symmetries of these $p$-form fields imply that all contributions to $\theta + \bar \theta$ in the IR, and thus the neutron EDM, are calculable from the dynamics of the theory.  Given that all models beyond the SM should be embedded in quantum gravity, this then leads us to question whether the entire discussion of global $P$ and $CP$ symmetries and their implications in~\cite{Kaplan:2025bgy} is not in the swampland.


\section*{\textbf{Acknowledgements}}
We thank Quentin Bonnefoy, Itay Bloch, Nathaniel Craig, Gian Francesco Giudice, Lawrence Hall, Daniel Harlow,
David B. Kaplan, David E. Kaplan, Tom Melia, Matthew McCullough, Jacob McNamara, Surjeet Rajendran, Maria Ramos, Matthew Reece, Mario Reig, Nicholas Rodd, Carlos Tamarit, David Tong, Gabriele Veneziano and Alexander Zhiboedov for useful discussions. We thank the Pollica Physics Center and CERN for hospitality during the writing of this manuscript. J.B. and B.R.S. are supported in part by the DOE award DESC0025293. C.A.M. is supported by the U.S. Department of Energy (DE-SC0009988) and the Sivian Fund.  AH is supported by NSF grant PHY-2514660 and the Maryland Center for Fundamental Physics.

\bibliography{Bibliography}

\clearpage
\appendix

\section{QED in 1+1D}
\label{app:1+1QED}

Here we discuss QED in 1+1D; which is an exactly solvable QFT that contains a $\theta$ term in the Lagrangian. This allows us to study the vacuum of this theory, which is in some ways analogous to that of QCD, by direct computation. We begin by discussing the vacuum structure after fully gauge fixing, and then proceed to study $\theta$ as a Lagrangian parameter as well as the associated boundary condition phase, via both canonical quantization  and the path integral. We also draw an analogy between the Strong CP problem and the the presence of an electric dipole moment of an $e^+e^-$ bound state that arises when massive fermions are included. Lastly we discuss gauging parity in this theory and show by explicit calculation that the partition function has contributions from orientable and non-orientable manifolds. 

\setcounter{equation}{0}

\subsection{Perspectives on theta vacua}

The Lagrangian density is the familiar one
\begin{align}
	{\cal L} &= - \frac{1}{4 e^2} F^{\mu\nu} F_{\mu\nu}	
	+ \frac{\theta_0}{4\pi} \epsilon_{\mu\nu} F^{\mu\nu} .
\end{align}
We consider the space to be a circle $x \in [0,L]$ where $x=0$ and $x=L$ are identified. 
We can always choose the Weyl (temporal) gauge $A^0 =0$ using the gauge parameter
\begin{align}
	\omega(x,t) = \int_0^t A^0(x,t') d t',
\end{align}
and perform a gauge transformation
\begin{align}
	A^0(x,t) & \rightarrow A^0 (x,t) - \partial_t \omega(x,t), \\
	A^1(x,t) & \rightarrow A^1 (x,t) + \partial_x \omega(x,t) .
\end{align}
By definition, $A^0(x,t)$ is transformed to zero, while $A^1$ is not. Now that $A^0 = 0$, the Euler--Lagrange equation with respect to the variation of $A^0$ becomes a constraint (namely, Gauss's law),
\begin{align}
	\partial_x E_x = - \partial_x \dot{A}^1 = 0.
	\label{eq:Gauss}
\end{align}
Here and below, we use the simplified notation $A = A^1$. The Lagrangian density in this gauge is
\begin{align}
	{\cal L} = \frac{1}{2e^2} \dot{A}^2  + \frac{\theta_0}{2\pi} \dot{A}.
\end{align}
Now $A$ is the dynamical variable in the theory and the canonical quantization condition is that
\begin{align}
	E_x = - \dot{A}, \qquad
	\left[A(x,t), {1 \over e^2} \dot{A}(y,t)\right] = i \delta(x-y) \,,
\end{align}
and its wave functions $\psi[A(x)] = \langle A(x) | \psi \rangle$ are subject to the ``physical state condition''
\begin{align}
	\partial_x E_x |{\rm phys}\rangle = 0.
\end{align}
It is easy to see that this condition is the gauge invariance of the state under ``small'' gauge transformation
\begin{align}
	\int d x&\, \omega(x) \langle A(x) | \partial_x E_x |{\rm phys}\rangle  \notag \\
    &= 
	\int d x\, \omega(x) \partial_x i \frac{\partial}{\partial A(x)} \langle A(x) | {\rm phys}\rangle 
	\nonumber \\
	&= -i\int d x\, \partial_x \omega(x)   \frac{\partial}{\partial A(x)} \langle A(x) | {\rm phys}\rangle 
	= 0.
	\label{eq:physical}
\end{align}
To make it clear what the physical state condition means, we expand the vector potential at a fixed time slice as
\begin{align}
	A(x) &= a_0 + \sum_{n=1}^{\infty} \left( a_n e^{2\pi i n x/L} + a_n ^*e^{-2\pi i n x/L} \right),
\end{align}
where $a_0$ is real and $a_n (n>1)$ are complex. The wave functional $\psi[A] = \langle A(x) | {\rm phys}\rangle$ is in principle a function of infinite number of variables $a_0, a_n, a_n^*$. However, the physical state condition ~\eqref{eq:physical} means that it does not depend on any $a_n, a_n^* (n >0)$. It is a function of $a_0$ only, $\psi[A] = \psi(a_0)$.

If we consider a ``fully gauge fixed" space of all gauge fields, we consider the wave functional to live on the space ${\cal A} / {\cal G}$, where ${\cal G}$ is a group of all possible gauge transformations on $S^1$. Keeping only $a_0$ eliminates redundancies up to gauge transformations that are continuously connected to the identity. However, there are also ``large'' gauge transformations $g_n(x)=e^{-2 \pi i n x/L}$
\begin{align}
	A(x) \rightarrow A(x) + i g_n^{-1}(x) \partial_x g_n(x)
	= A(x) + \frac{2\pi}{L} n.
\end{align}
It is clear that $g_n$ winds U(1) $n$ times as we move along $x \in S^1$. $n$ is the winding number
\begin{align}
	n &= \frac{i}{2\pi} \int_0^L d x g_n^{-1}(x) \partial_x g_n(x).
\end{align}
Therefore, ${\cal G} = \oplus_n {\cal G}_n$ where ${\cal G}_n$ is a set of all possible gauge transformations with winding number $n$.  $a_0 \in {\mathbb R}$ is the coordinate of ${\cal A}/{\cal G}_0$, while $a_0 \in [0, \frac{2\pi}{L}]$ is the coordinate of ${\cal A}/{\cal G}$. Because of the identification by ${\cal G}$, the space ${\cal A} / {\cal G}$ has a non-contractible loop. The wave functional can have a non-zero phase when the gauge field goes around this loop. This phase is not determined by the action.

The fact that $a_0$ is defined only modulo $2\pi/L$ is evident when we look at gauge-invariant observable such as the Wilson loop in the spatial direction,
\begin{align}
	W = e^{i \oint A_1 d x} = e^{i a_0 L}.
\end{align}
It is clear that changing $a_0 \rightarrow a_0 + \frac{2\pi}{L} n$ does not change the Wilson loop.

The ``fully gauge fixed'' $a_0$ lives on a one-dimensional circle, and the theory is identical to that of a quantum particle on 1D circle with the action
\begin{align}
	S &= \int dt dx {\cal L}
	= L \int dt \left[ \frac{1}{2e^2} \dot{a}_0^2 + \frac{\theta_0}{2\pi} \dot{a}_0 \right].
\end{align}
It is clear that $\theta_0$ is analogous to the Aharonov--Bohm phase. 
On the other hand, just like the quantum particle on a 1D circle, there is also an arbitrariness in gluing the wave function where $a_0 = \frac{2\pi}{L}$ (see, {\it e.g.}, the discussions in~\cite{Bonneau:1999zq,TongGauge}),\footnote{
The general boundary condition for a particle on a circle given in~\eqref{eq:psi} can be understood through the following argument. Quantum operators should be self-adjoint rather than merely symmetric in order to give real values of observables and preserve unitarity under time evolution. However, some symmetry operators admit multiple self-adjoint extensions, which are quantified through the deficiency index, and which are represented by different, independent boundary conditions that preserve the self-adjoint requirement. Consider the momentum operator $\hat P = -i \partial_x$ in quantum mechanics on a circle of circumference $L$, with two states in the Hilbert space $| \psi \rangle$ and $|\chi \rangle$. The self-adjoint requirement enforces $\langle \psi | \hat P \chi \rangle = \langle \hat P \psi | \chi \rangle$, which in position space implies
\es{}{
\int_0^L dx\, \left( \bar \psi \partial_x \chi + \partial_x \bar \psi \chi \right) = 0 \,.
}
Integrating by parts the equation above trivially vanishes, except for the boundary contributions, giving the self-adjoint requirement 
\es{}{
\chi(L) \bar \psi(L) = \chi(0) \bar \psi(0) \,.
}
This condition must be satisfied for all states in the domain.
This requirement is not just satisfied by the normal periodic boundary condition but, more generally, by the boundary conditions $\psi(L) = e^{i \alpha}\psi(0)$ for any choice of $\alpha$, so long as this boundary condition is applied consistently to all states.  The choice of boundary condition can be thought of as a choice of domain (or a choice of superselection sector) that is needed to make the momentum operator well defined on this compact space~\cite{Bonneau:1999zq}.
}
\begin{align}
	\psi\left( \frac{2\pi}{L} \right) = e^{i\theta} \psi(0).
	\label{eq:psi}
\end{align}
Physical observables depend only on the combination $\theta_0  - \theta$.  One way of understanding this is to consider expectation values of the Hamiltonian, which in this case, as we discuss further below, is a function of the canonical momentum operator $\Pi = -i \partial_{a_0}$ in the specific combination $\Pi - {L \theta_0 \over 2\pi}$.  The boundary condition in~\eqref{eq:psi} implies that eigenstates of the momentum operator have eigenvalues $\Pi_n = \theta L / (2 \pi) + n L$, for integer $n \geq 0$, which implies that eigenstates of the Hamiltonian only depend on the combination $\theta - \theta_0$. 

\begin{figure}[t]
\centerline{
\includegraphics[width=0.5\textwidth]{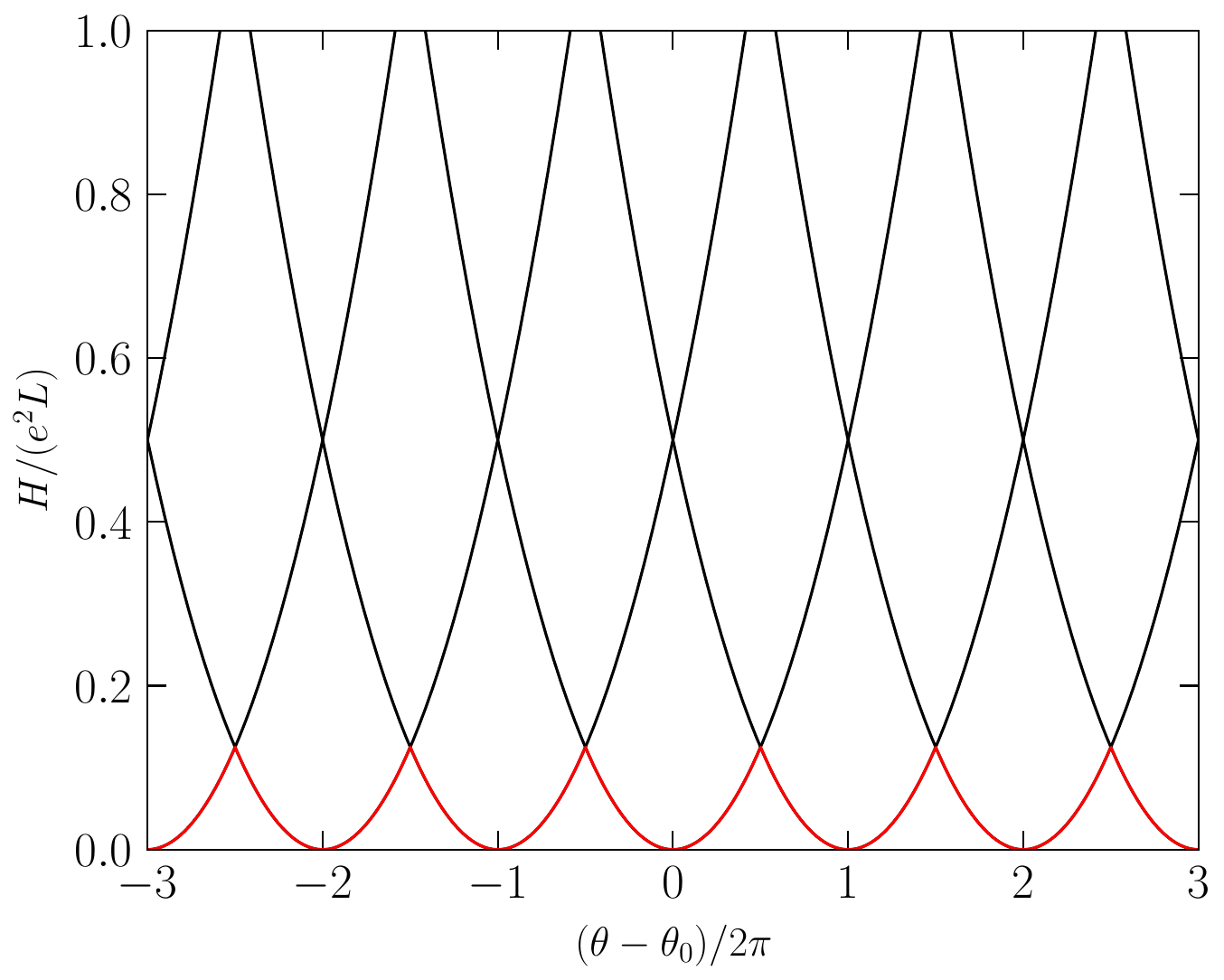}
}
\caption{The eigenvalues of the Hamiltonian of  1+1D QED on $S^{1}$ as a function of the  angle $\theta - \theta_0$ (black) with the ground state indicated in red.} \label{fig:eigenvalues}
\end{figure}

Below, we compute the transition amplitude $\langle a_f, t_f | a_i, t_i \rangle$ both with the canonical formulation of quantum field theory and path integral formulation to understand the consistency as well as how each formulation handles the Lagrangian parameter and the boundary condition.  These results reinforce the conclusion that physical observables only depend on the combination $\theta - \theta_0$ and cannot distinguish the two contributions separately. 

Before turning to these calculations, however, let us comment briefly on how to generalize the lessons from this example to QCD in 3+1D and the extra-dimensional constructions, for example in 5D, that we discuss in the main text.  First, let us understand the following conundrum. To an observer living in 1+1D on a circle, we show above that there is a non-unique way of defining the periodic boundary conditions for wavefunctions, characterized by the angle $\theta$ as given in~\eqref{eq:psi}.  Different values of $\theta$ lead to physically distinct observables, such as spectra, though $\theta$ is only observable in the combination $\theta - \theta_0$, with $\theta_0$ the Lagrangian term. Now let us imagine that our 1+1D example arises from the dimensional reduction of QED in 3+1D for particles that are localized on a circle. In 3+1D Minkowski space the deficiency index for {\it e.g.} the momentum operator is trivial, meaning that there is no ambiguity in defining boundary conditions for wavefunctions in QED. How, then, do we understand where the combination $\theta - \theta_0$ arises from the UV perspective? The answer is that through the Aharonov-Bohm effect the combination $\theta - \theta_0$ is proportional to the magnetic flux through the circle on which our 1+1D QED lives. The 1+1D observable is not able to access the interior of this circle and so is unaware of the concept of magnetic flux; to this observer, $\theta_0$ is simply a Lagrangian parameter allowed by the theory and $\theta$ is a generalized boundary condition that is allowed by quantum mechanics. But to the 3+1D observer who has access to the UV theory, the physical combination $\theta - \theta_0$ is simply set by the magnetic flux through the circle.

Let us now apply this analogy to the theory of QCD in 3+1D and a UV completion that arises from the dimensional reduction of a 4+1D theory on a circle.  Here, we want to ask how the wave-functional $\psi(A')$ is related to $\psi(A)$ for $A'$ that are given by gauge transformations of the QCD vector potential $A$ (here, our discussion mirrors that of~\cite{TongGauge}). For $A'$ that are given by gauge transformations that are continuously deformable to the identity ({\it i.e.}, topologically trivial gauge transformations), we must take $\psi(A') = \psi(A)$. On the other hand, for topologically non-trivial gauge transformations, which are characterized by a winding number $n$, we are allowed to take $\psi(A') = e^{-i \theta n} \psi(A)$, in analogy to the 1+1D QED example discussed above and the quantum mechanics of a particle on a ring (see~\cite{TongGauge}). This choice of domain is precisely the choice of superselection sector in the theta vacua, but just as in 1+1D QED $\theta$ only enters in combination with the Lagrangian parameter $\theta_0$ for physical observables.  

Now, let us consider our 5D example, where we have non-abelian gauge theory in 5D that descends to QCD in 4D by compactification on an $S^1$ of size $R$. At high energies ($E \gg R^{-1}$) spacetime looks locally like ${\mathbb R}^{4,1}$, so that the vacuum of the theory is characterized by $\pi_4(SU(N))$, which is trivial for $N \geq 3$. Thus, there is no ambiguity in the UV in terms of the structure of the vacuum and also, as we have discussed, there is no theta term in the Lagrangian in 5D. On the other hand, the 5D theory does have a 5D CS term of the form $A \wedge F \wedge F$, and under dimensional reduction to 4D this can generate a theta term in the Lagrangian if there are non-trivial fluxes across the $S^1$.  In this case, the combination $\theta + \bar \theta$ in the IR EFT in 4D is precisely given by the magnetic fluxes of the 5D theory through the $S^1$. This is in direct analogy to our example discussed above of the dimensional reduction of QED in 4D to QED in 2D. The 4D theory has a non-ambiguous vacuum, while the 2D theory has an ambiguity in defining the vacuum along with a theta term---the combination of the two is physical. Through the Aharonov-Bohm effect the combination of these two parameters in 1+1D is set by the magnetic flux through the spatial circle.
\footnote{As a side-note, a crucial part of the arguments above is the compact nature of space, for example the compact circle in the 1+1D example.  In defining the winding numbers in QCD we also implicitly compactify space at infinity.  In contrast, the axial gauge introduced in Refs.~\cite{Bernard:1976wt,Weinberg:1992hc} cannot be defined on a compact space; it must be formally infinite \cite{Jackiw:1983nv}. In this case, unsurprisingly, there is no ambiguity in defining the wavefunction ({\it i.e.}, no superselection sectors for QCD). This result seems in contrast with the standard derivation in temporal gauge. We could not solve this puzzle. On the other hand, the Lagrangian still has a theta parameter.  In UV complete examples where $\theta + \bar \theta$ is calculable there is no contradiction, since it is the combination $\theta + \bar \theta$ that is set by the dynamics in the UV.   } 

\subsection{Canonical and Path Integral Formulation of QED in 1+1D}

Let us now return to the 1+1D QED example discussed above to show explicitly that amplitudes depend on the combination $\theta - \theta_0$, both for the canonical formulation of quantum field theory and the path integral formulation.
Starting with the canonical formulation, from the action we derive the canonical momentum operator
\es{}{
\hat \pi = L \left(\frac{1}{e^2} \dot{a}_0 + \frac{\theta_0}{2\pi} \right).
}
Solving for $\dot{a}_0$, we find
\begin{align}
	\dot{a}_0 &= \frac{e^2}{L} \hat \pi - \frac{e^2 \theta_0}{2\pi}.
\end{align}
The Hamiltonian is 
\begin{align}
	H &= \hat \pi \dot{a}_0  - L \left[ \frac{1}{2e^2} \dot{a}_0^2 + \frac{\theta_0}{2\pi} \dot{a}_0 \right]
    = 
    \frac{e^2}{2L} \left( \hat \pi - \frac{L \theta_0}{2\pi}\right)^2.
\end{align}
As usual, the canonical momentum $\hat \pi$ is represented as
\begin{align}
	\hat \pi = -i \frac{\partial}{\partial a_0} \ .
\end{align}
A general solution to the boundary condition ~\eqref{eq:psi} is
\begin{align}
	\psi_n (a_0) = \left( \frac{L}{2\pi} \right)^{1/2} 
	e^{i \left(n + \frac{\theta}{2\pi}\right) a_0L},
	\label{eq:eigenstates}
\end{align}
such that
\begin{align}
	\psi_n \left(a_0 + \frac{2\pi}{L} \right) 
	& = \left( \frac{L}{2\pi} \right)^{1/2} 
	e^{i \left(n + \frac{\theta}{2\pi}\right) \left(a_0 + \frac{2\pi}{L} \right)L}
	= e^{i\theta} \psi_n (a_0).
\end{align}
Therefore, the energy eigenvalues are
\begin{align}
	H \psi_n 
	= \frac{e^2L}{2} \left( n - \frac{\theta_0 - \theta}{2\pi} \right)^2 \psi_n .
\end{align}
The Wilson loop operator takes one eigenstate to another,
\begin{align}
	W \psi_n (a_0) &= e^{i a_0 L} \left( \frac{L}{2\pi} \right)^{1/2} 
	e^{i \left(n + \frac{\theta}{2\pi}\right) a_0L} \nonumber \\
	&= \left( \frac{L}{2\pi} \right)^{1/2} 
	e^{i \left(n + 1 + \frac{\theta}{2\pi}\right) a_0L}  = \psi_{n+1} (a_0).
\end{align}
On the other hand, a ``large gauge transformation'' is given by the unitarity operator
\begin{align}
	U = e^{2\pi i \hat \pi /L} .
\end{align}
It is easy to see
\begin{align}
	U \psi_n (a_0) = e^{2\pi i \left(n + \frac{\theta}{2\pi} \right)} \psi_n (a_0)
	= e^{i\theta} \psi_n (a_0).
\end{align}

The transition amplitude can be obtained readily as
\begin{align}
	\lefteqn{
	\langle a_f, t_f | a_i, t_i \rangle
	} \nonumber \\
	&= \sum_n \psi_n (a_f) e^{- i E_n (t_f - t_i)} \psi^*_n (a_i) \nonumber \\
	&=  \frac{L}{2\pi} \sum_n e^{i \left(n + \frac{\theta}{2\pi}\right) (a_f-a_i)  L} 
	e^{- i \frac{e^2L}{2} \left( n - \frac{\theta_0 - \theta}{2\pi}\right)^2 (t_f - t_i)} .
\end{align}
Using the notation $\Delta \theta = \theta_0 - \theta$, we obtain
\begin{align}
	\langle a_f, t_f | a_i, t_i \rangle
	&=  \frac{L}{2\pi} \sum_n e^{i \left(n + \frac{\theta}{2\pi}\right) (a_f-a_i)  L} 
	e^{- i \frac{e^2L}{2} \left( n - \frac{\Delta\theta}{2\pi}\right)^2 (t_f - t_i)} .
\end{align}
Using $T=t_f-t_i$, we obtain
\begin{widetext}
\begin{align}
	\langle a_f, t_f | a_i, t_i \rangle
	&=  \frac{L}{2\pi} \sum_n 
	e^{- i \frac{e^2 LT}{2}  n^2 
	+i \left((a_f-a_i)  L+  e^2LT\frac{\Delta\theta}{2\pi}\right) n
	+i  \frac{\theta}{2\pi}(a_f-a_i)  L
	- i \frac{e^2LT}{2} \left( \frac{\Delta\theta}{2\pi} \right)^2} 
	\nonumber\\
	&= \frac{L}{2\pi} e^{i  \frac{\theta}{2\pi}(a_f-a_i)  L
	- i \frac{e^2LT}{2} \left( \frac{\Delta\theta}{2\pi} \right)^2} 
	\vartheta(z;\tau),
	\label{eq:transition}
\end{align}
\end{widetext}
where $\vartheta(z;\tau)$ is a Jacobi theta function
\begin{align}
	\vartheta(z;\tau) \equiv \sum_{n=\infty}^{+\infty} q^{n^2} \eta^n
	= \sum_{n=\infty}^{+\infty} e^{\pi i n^2 \tau + 2 \pi i n z}\ .
\end{align}
Therefore for our case,
\begin{align}
	z &= 	\frac{1}{2\pi} \left((a_f-a_i)  L +  e^2LT\frac{\Delta\theta}{2\pi}\right) , \\
	\tau &= -  \frac{e^2 LT}{2\pi}  .
\end{align}

We can also derive the path integral from ~\eqref{eq:transition}. We need to work out its small $T$ behavior. Using the modular transformation, 
\begin{align}
	\vartheta\left(\frac{z}{\tau}; \frac{-1}{\tau} \right)
	&= \alpha \vartheta(z; \tau), \\
	\alpha &= (-i\tau)^{1/2} e^{\pi i z^2/\tau},
\end{align}
we obtain
\begin{widetext}
\begin{align}
	\langle a_f, t_f | a_i, t_i \rangle
	&= \frac{L}{2\pi} e^{i  \frac{\theta}{2\pi}(a_f-a_i)  L
	- i \frac{e^2LT}{2} \left( \frac{\Delta\theta}{2\pi} \right)^2} 
	(-i\tau)^{-1/2} e^{-\pi i z^2/\tau}
	\vartheta\left(\frac{z}{\tau}; \frac{-1}{\tau}\right) \nonumber \\
	&= 
	\left(\frac{L}{2\pi i e^2T}\right)^{1/2} 
	e^{ i\frac{(a_f-a_i)^2  L}{2e^2 T} 
	+   i (a_f-a_i)  L \frac{\theta_0}{2\pi} }
	\vartheta\left(\frac{z}{\tau}; \frac{-1}{\tau}\right) .
	\label{eq:transition2}
\end{align}
\end{widetext}
For small $\tau$ in the upper half plane, the new $q'=e^{-\pi i /\tau} \ll 1$ and hence
\begin{align}
	\vartheta\left(\frac{z}{\tau}; \frac{-1}{\tau}\right) 
	= 1 + O(q').
\end{align}
We find
\begin{align}
	\langle a_f, t_f | a_i, t_i \rangle
	&\simeq  
	\left(\frac{L}{2\pi i e^2T}\right)^{1/2} 
	e^{ i\frac{(a_f-a_i)^2  L}{2e^2 T} 
	+   i (a_f-a_i)  L \frac{\theta_0}{2\pi} }
	\label{eq:smallT}
\end{align}
for small $T$. Therefore, the path integral is given by
\begin{align}
	\langle a_f, t_f | a_i, t_i \rangle
	&= \int {\cal D}a(t) e^{ i \int_{t_i}^{t_f} d t \left[\frac{\dot{a}^2  L}{2e^2} 
	+   \dot{a}  L \frac{\theta_0}{2\pi} \right]}\ ,
\end{align}
where the singular prefactor in ~\eqref{eq:smallT} is included into the path integral measure. Finally, in order to make the path integral self-contained, we need to remove the boundary condition phase $e^{i\theta}$ from the initial and final states. Then we obtain the expression
\begin{align}
	\langle a_f, t_f | a_i, t_i \rangle
	&= e^{-i (a_f - a_i) L \frac{\theta}{2\pi}} 
	\int {\cal D}a(t) e^{ i \int_{t_i}^{t_f} d t \left[\frac{\dot{a}^2  L}{2e^2} 
	+  \dot{a}  L \frac{\theta_0}{2\pi} \right]} \nonumber \\
	& = \int {\cal D}a(t) e^{ i \int_{t_i}^{t_f} d t \left[\frac{\dot{a}^2  L}{2e^2} 
	+  \dot{a}  L \frac{\Delta \theta}{2\pi} \right]}\,.
\end{align}
Thus, both in the canonical and path integral formulations, amplitudes only depend on $\Delta \theta$.

\subsection{An analogy to the Strong  $CP$ Problem}
We may also identify a phenomenon in this theory analogous to the dependence of the neutron electric dipole moment on the QCD vacuum angle. Let us now consider 1+1D QED coupled to massive matter particles of charge $e$ which we call electrons. If $m \gg e$, the electron is non-relativistic. Suppose there is a pair of an electron and a positron. Their Lagrangian is
\begin{align}
L&=\frac{1}{2} m\left(\dot{x}_1^2+\dot{x}_2^2\right)-A^0\left(x_1\right)+a_0\left(x_1\right) \dot{x}_1 \notag \\
&+A^0\left(x_2\right)-a_0\left(x_2\right) \dot{x}_2 \,.
\end{align}
Then the Gauss' law constraint is modified to
$$
\partial_x E=-\partial_x^2 A^0-\partial_x \dot{a}_0=e^2 \delta\left(x-x_1(t)\right)-e^2 \delta\left(x-x_2(t)\right),
$$
where $x_1$ is the position of the positron and $x_2$ that of electron. The solution is

$$
\begin{aligned}
A^0(x)&=e^2\left\{\begin{array}{ll}
\frac{1}{L}\left(x_2-x_1\right) x & 0 \leq x \leq x_1 \\
-\left(x-x_1\right)+\frac{1}{L}\left(x_2-x_1\right) x & x_1 \leq x \leq x_2 \\
-\left(x_2-x_1\right)+\frac{1}{L}\left(x_2-x_1\right) x & x_2 \leq x \leq L
\end{array},\right. \\
E(x)&=-\dot{a}_0+e^2 \begin{cases}-\frac{1}{L}\left(x_2-x_1\right) & 0 \leq x \leq x_1 \\
1-\frac{1}{L}\left(x_2-x_1\right) & x_1 \leq x \leq x_2 \\
-\frac{1}{L}\left(x_2-x_1\right) & x_2 \leq x \leq L\end{cases}
\end{aligned}
$$
where we assume $x_1<x_2$. When $x_1>x_2$,

$$
\begin{gathered}
A^0(x)=e^2 \begin{cases}-\frac{1}{L}\left(x_1-x_2\right) x & 0 \leq x \leq x_2 \\
\left(x-x_2\right)-\frac{1}{L}\left(x_1-x_2\right) x & x_2 \leq x \leq x_1 \\
\left(x_1-x_2\right)-\frac{1}{L}\left(x_1-x_2\right) x & x_1 \leq x \leq L\end{cases} \\
E(x)=-\dot{a}_0+e^2 \begin{cases}\frac{1}{L}\left(x_1-x_2\right) & 0 \leq x \leq x_1 \\
-1+\frac{1}{L}\left(x_1-x_2\right) & x_2 \leq x \leq x_1 \\
\frac{1}{L}\left(x_1-x_2\right) & x_1 \leq x \leq L\end{cases} \,.
\end{gathered}
$$
The constant piece of the electric field depends on the wave function \eqref{eq:eigenstates},
\begin{align}
E \psi_n\left(a_0\right)&=-\dot{a}_0 \psi_n\left(a_0\right)\notag \\
&=-e^2\left(n-\frac{\theta_0-\theta}{2 \pi}\right) \psi_n\left(a_0\right)
=E_n \psi_n\left(a_0\right) ;
\end{align}
that is, the ground state energy is 
\es{}{
E_0 = e^2 \left( {\theta_0 - \theta \over 2 \pi} \right) \,.
}

In the ground state, $-\frac{1}{2} e^2<E_0<\frac{1}{2} e^2$.
We are interested in the limit $\left|x_1-x_2\right| \ll L \rightarrow \infty$. Then

$$
\int \frac{1}{2 e^2} E^2 d x=\frac{1}{2 e^2} E_0^2 L+E_0\left(x_2-x_1\right)+\frac{1}{2} e^2\left|x_2-x_1\right| \,.
$$
Therefore, the electron and positron have the Hamiltonian
$$
H=\frac{1}{2 m}\left(p_1^2+p_2^2\right)+\frac{1}{2} e^2\left|x_2-x_1\right|+E_0\left(x_2-x_1\right) .
$$
After the center of mass motion is separated, we obtain the Hamiltonian for the relative motion

$$
H=\frac{1}{2 \mu} p_1^2+\frac{1}{2} e^2|x|+E_0 x \,,
$$
with $\mu=\frac{1}{2} m$. With the term $|x|$, the electron and positron are confined. Yet when $E_0=\frac{e^2}{2}\left(\theta_0-\theta=\pi, n=0\right)$, it is deconfined in the $x \rightarrow-\infty$ direction while when $E_0=-\frac{e^2}{2}\left(\theta_0-\theta=-\pi, n=0\right)$, it is deconfined in the $x \rightarrow+\infty$ direction. More generally, $E_0>0$ makes the confining potential steeper in positive $x$ and shallower in negative $x$ direction, and therefore the center of the wave function is shifted to a negative value. It exhibits an electric dipole moment $d=\langle-e x\rangle \neq 0$.

\subsection{Gauging parity in 1+1D QED}
\label{app:gauge_parity_1+1QED}

QED on the torus $T^2$ is nothing but the partition function
\begin{align}
	Z = {\rm Tr} e^{-\beta H}
	= \int_0^{2\pi/L} d a \langle a, t_f | a, t_i \rangle,
\end{align}
where $-\beta = -i (t_f - t_i)$. 
We can also consider QED on the Klein bottle $K^2$, which is a non-orientable two-dimensional surface, where two ends of a cylinder are identified after reversing the orientation. Namely that it is a torus with a parity twist. Therefore, we regard QED on $K^2$ as a theory on $S^1$ but take the partition function with parity operator
\begin{align}
	Z_{-} = {\rm Tr} P e^{-\beta H}.
\end{align}
Then the total partition function with the gauged parity has a projection operator
\begin{align}
	Z = Z_+ + Z_-
	= {\rm Tr} (1+P) e^{-\beta H}.
\end{align}
Note that this result is analogous to the  one-loop closed-string contributions to the partition function of the Type I string worldsheet theory discussed in \ref{app:typeI}.

Since the parity operation can be done at any point along the time axis with two patches overlapping, the parity operator needs to commute with the Hamiltonian. Namely that the Hamiltonian must be parity invariant. We first take the Lagrangian parameter $\theta_0=0$. We will come back to the other possible case of $\theta_0 = \pi$ later. The Hamiltonian eigenstates are ~\eqref{eq:eigenstates},
\begin{align}
	\psi_n (a_0) = \left( \frac{L}{2\pi} \right)^{1/2} 
	e^{i \left(n + \frac{\theta}{2\pi}\right) a_0L}.
\end{align}
To require the states survive under the projection $1+P$ within the same Hilbert space, we must either take $\theta=0$ and have the Hamiltonian eigenstates
\begin{align}
	\phi_0 (a_0) &= \left( \frac{L}{2\pi} \right)^{1/2}, \\
	\phi_n (a_0) &= \left( \frac{L}{\pi} \right)^{1/2} \cos (n a_0L) \quad (n \geq 1),
\end{align}
or take $\theta=\pi$ and
\begin{align}
	\phi_n (a_0) = \left( \frac{L}{\pi} \right)^{1/2} \cos \left(n + \frac{1}{2} \right) a_0L.
\end{align}
The sum is now reduced over the subsets.

For the Hilbert space with $\theta = 0$, the transition matrix element is
\begin{widetext}
\begin{align}
	\langle a_f, t_f | a_i, t_i \rangle
	&=  \frac{L}{2\pi} \left[ 1 + 2 \sum_{n =1}^\infty
	e^{- i \frac{e^2 LT}{2}  n^2 } 
	\cos (n a_i L) \cos (n a_f L) \right]
	\nonumber\\
	&=  \frac{L}{2\pi} \left[ 1 +  \frac{1}{2} \sum_{n =1}^\infty
	e^{- i \frac{e^2 LT}{2}  n^2 } 
	(e^{i n (a_f - a_i) L} + e^{i n (a_f + a_i) L}
	+ e^{-i n (a_f - a_i) L} + e^{-i n (a_f + a_i) L} ) \right]
	\nonumber\\
	&=  \frac{L}{2\pi}  \frac{1}{2} \sum_{n}
	e^{- i \frac{e^2 LT}{2}  n^2 } 
	(e^{-i n (a_f - a_i) L} + e^{-i n (a_f + a_i) L} )
	\nonumber\\
	&= \frac{L}{2\pi} \frac{1}{2}
	(\vartheta(z_1;\tau) + \vartheta(z_2;\tau) ),
\end{align}
\end{widetext}
where
\begin{align}
	z_1 &= \frac{1}{2\pi} (a_f-a_i)  L  , \\
	z_2 &= \frac{1}{2\pi} (a_f+a_i)  L  , \\
	\tau &= -  \frac{e^2 LT}{2\pi}  .
\end{align}
For the Hilbert space with $\theta = \pi$, a similar calculation shows the transition matrix element is
\begin{align}
	\langle a_f, t_f | a_i, t_i \rangle
	&=  
    \frac{L}{2\pi} \frac{1}{2}
	(\vartheta_{10}(z_1;\tau) + \vartheta_{10}(z_2;\tau) ).
\end{align}
For both cases, the result is indeed the average of that for $a_f$ and $-a_f$ as expected from the operator $1+P$ inserted. 

\section{Gauged worldsheet parity in Type-I string theory}
\label{app:typeI}

It is instructive to discuss an analogous example of a gauged spacetime inversion that is commonplace in string theory: worldsheet-parity. The embedding of a fundamental string in spacetime is described by the  worldsheet $X^\mu(\sigma^0,\sigma^1)$ with $\sigma^1$ the space-like coordinate along the length of the string and $\sigma^0$ the time-like coordinate along the string's evolution. Type IIB string theory is a theory of closed, oriented  strings which is left-right symmetric. That is, it is symmetric under the worldsheet parity transformation $\Omega: (\sigma^0, \sigma^1) \to (\sigma^0,2\pi-\sigma^1)$, which exchanges left and right movers. Type I string theory is a theory of unoriented strings, and is obtained by modding out Type IIB by $\Omega$ (``orientifolding"), meaning that states related by $\Omega$, e.g,
\begin{equation}
|a\rangle_L \otimes|b\rangle_R \,,\quad|a\rangle_R \otimes|b\rangle_R
\end{equation}
are considered equivalent, so in Type I $\Omega$ is a gauge symmetry. In addition to the closed unoriented strings,  an open string sector is required for tadpole and anomaly cancellation. Gauging $\Omega$ halves the Hilbert space, projecting out states which are not invariant under $\Omega$ ({\it e.g.}, the open string photon and the antisymmetric tensor $B_{\mu\nu}$, which are $\Omega$-odd; the dilaton and graviton survive).  In string perturbation theory, string S-matrix elements are computed via the Polyakov path integral, in which the sum over metrics and worldsheet topologies is organized by power counting in the string coupling $g_s$, with the power set by the Euler characteristic $\chi$ of the worldsheet topology:
\begin{equation}
Z=\sum_{\substack{\text { topologies } \\ \text { metrics }}} e^{-S_{\text {string }}} \sim \sum_{\text {topologies }} g_s^{-\chi} \int \mathcal{D} X \mathcal{D} g e^{-S_{\text {Poly }}} \,,
\end{equation}
with $Z$ the partition function,  and
\begin{equation}
S_{\text {Poly }}=\frac{1}{2} T \int_{\mathcal{M}} \mathrm{d}\sigma^0\mathrm{d}\sigma^1 \sqrt{g} g^{\alpha \beta}  \partial_\alpha X^\mu \partial_\beta X_\nu
\end{equation}
is the Polyakov action (here $T=1/(2\pi\alpha')$ is the string tension). The integration over worldsheet spacetime coordinates $X$ and worldsheet metrics $g$ is modulo diffeomorphisms and Weyl transformations (note that the target spacetime metric is fixed).
For fixed $\chi$ there are a finite number of worldsheet topologies which contribute, so the sum over topologies is perfectly well-defined and at low orders S-matrix elements can be easily computed. For example, in Type IIB the leading order ($\chi=2$) diagram is $S^2$, and the ``one-loop" ($\chi=0$) contribution comes from the torus $T^2$. 

In Type I, the path integral is modified to include contributions from both orientable and non-orientable worldsheets.  Concretely, in the unoriented theory the one-loop vacuum amplitude contribution from $T^2$ encodes closed strings evolving and gluing back to their original state. The torus may be parametrized in the complex plane by the modulus $\tau=\tau_1+i\tau_2$ with identifications of the complex coordinate $\omega$, $\omega \simeq \omega+2\pi \simeq \omega+2\pi\tau$. The worldsheet theory in this setup is constructed by considering a field theory on a spatial circle with coordinate $\sigma^1=\mathrm{Re}(\omega)$, and then evolving $\sigma^2=\mathrm{Im}(\omega)$ for Euclidean time interval of length $2\pi\tau_2$, while also shifting $\sigma^1$ by $2\pi \tau_1$, and then identifying the ends. The partition function is therefore a trace over the Hilbert space $\mathcal{H}_c$ of the closed string sector, which in light-cone gauge takes the form \cite{Burgess:1986ah,Polchinski:1998rq,Polchinski:1998rr,Angelantonj:2002ct,Ibanez:2012zz,Polchinski:1994mb,Staessens:2010vi}
\begin{equation}
Z_{T^2}=\operatorname{Tr}_{\mathcal{H}_c}\left[e^{2\pi i\tau_1P-2\pi\tau_2H}\right]=\operatorname{Tr}_{\mathcal{H}_c}\left[q^{L_0 -\frac{c}{24}} \bar{q}^{\tilde{L}_0-\frac{\tilde{c}}{24}}\right] \,,
\end{equation}
with $q=e^{2\pi i\tau}$ and $L_0$ $(\tilde{L_0})$ are the left (right) moving Virasoro zero-mode operators and $c$ ($\tilde{c}$) are the left (right) moving central charges of the worldsheet conformal field theory. Above, the momentum $P=L_0-\tilde{L_0}$ and the Hamiltonian $H=L_0+\tilde{L_0}-\frac{(c+\tilde{c})}{24}$ generate translations along $\sigma^1$ and $\sigma^2$ respectively. Gauging $\Omega$ corresponds to inserting the projector $P=\frac{1}{2}(1+\Omega)$  
\begin{equation}
Z_\mathrm{closed}=\operatorname{Tr}_{\mathcal{H}_c}\left[P q^{L_0} \bar{q}^{\tilde{L}_0}\right]=\frac{1}{2}Z_{T^2} + Z_{K^2}\,,
\label{eq:Zclosed}
\end{equation}
such that there is a contribution from the Klein bottle $K^2$, $Z_{K^2} = \frac{1}{2} \operatorname{Tr}_{\mathcal{H}_c}\left[\Omega q^{L_0-\frac{c}{24}} \bar{q}^{\tilde{L}_0-\frac{\tilde{c}}{24}}\right]$, corresponding to closed strings evolving and gluing back to their original state up to the action of $\Omega$. The contributions in \eqref{eq:Zclosed} can be computed exactly (see {\it e.g.}, \cite{Polchinski:1998rq,Polchinski:1998rr}). This discussion is analogous to gauging parity in 1+1D QED, which we discuss in Appendix \ref{app:1+1QED}, where we explicitly compute the partition function of the theory on $T^2$ and $K^2$.   

To account for the open-string sector in Type I string theory, the sum over worldsheets must include topologies with boundary. At ``half-loop" order $\chi=1$ these are the disk $D^2$ and the real projective plane $\mathbb{R}P^2$, which is unorientable. At one loop these are the cylinder and the Möbius strip. The vacuum amplitude contributions between the cylinder and the unorientable Möbius strip are related by parity projection similarly to the torus and Klein bottle. It is natural to ask, given this pairing between orientable and unorientable worldsheets at one-loop order associated to gauging parity, whether a similar pairing happens at tree-level and half-loop order. This cannot happen because there are an odd number of topologies at these orders: $S^2$, $D^2$, and $\mathbb{R}P^2$. The key point is that these manifolds, unlike for those appearing at one-loop, do not have non-contractible loops, so there are no Euclidean time-circles that allow the partition function to be written as a trace over the Hilbert space. At tree-level and half-loop order, gauging worldsheet parity therefore leaves the partition function unchanged, though the Hilbert space is still projected down to $\Omega$-invariant states ({\it c.f.} our previous discussion on gauging spacetime parity in Sec.~\ref{sec:symmetry_solns_strongCP}).   

\section{The Witten-Veneziano relation}
\label{app:WV_derivation}
Here we recall how the Witten-Veneziano formula \eqref{eq:WV_main} for the mass of the $\eta'$ can be derived via large $N$ arguments following the original work by Witten \cite{Witten:1979vv}. We stress that this derivation does not assume anything about the order of limits in the partition function. 
We work in pure Yang-Mills, and assume the 't Hooft limit, with $\lambda=g^2N_c$ fixed, such that the Lagrangian takes the form
\begin{align}
\mathcal{L}_\mathrm{Y M} 
& =N_c\left(-\frac{1}{2 \lambda} \operatorname{tr} F_{\mu \nu} F^{\mu \nu}+\frac{\theta}{16 \pi^2 N_c} \operatorname{tr} F_{\mu \nu}\tilde{F}^{\mu \nu}\right) \,.
\label{eq:Lagrangian_largeN_YM}
\end{align}
The topological susceptibility is
\begin{equation}
\chi(k)=\int d^4 x e^{i k \cdot x}\left\langle q(x)q(0) \right\rangle \,,
\end{equation}
with $q(x)=F\tilde{F}$, such that
\begin{equation}
\frac{d^2 E}{d \theta^2}=\left(\frac{1}{16 \pi^2 N_c}\right)^2 \lim _{k \rightarrow 0} \chi(k) \,.
\end{equation}
To leading order, the singularities of two-point function of the topological charge density arise from glueballs and flavor-singlet pseudoscalar mesons,
\begin{equation}
\chi(k)=\sum_{\text {glueballs }} \frac{a_n^2}{k^2-M_n^2}+\sum_{\text {mesons }} \frac{b_n^2}{N(k^2-m_n^2)} \,,
\label{eq:chik}
\end{equation}
with $M_n$ ($m_n$) the glueball (meson) masses, and $a_n$ ($b_n$) the
amplitudes for $q$ to create these states from the vacuum.

If massless quarks are introduced, then the topological susceptibility must be zero, since a chiral rotation of the quarks can make $\theta$ disappear, such that it cannot have any physical effects. However, quark loops contribute to $\chi(k)$ only at order $1/N_c$, whereas the gluons contribute at order $1/N_c^0$. The cancellation that must occur in \eqref{eq:chik} at $k=0$ is between the leading order meson pole from the $\eta'$ and the glueball contribution (which is just the topological susceptibility of pure Yang-Mills $\chi_\mathrm{YM}$). Note that this can only happen if $m^2_\mathrm{\eta'}$ is of order $1/N$. This cancellation tells us that
\begin{equation}
\chi_\mathrm{YM} \equiv \left.\chi(0)\right|_{\text {Yang-Mills }}=\frac{b_{\eta^{\prime}}^2}{Nm_{\eta^{\prime}}^2} \,.
\end{equation}
The axial anomaly gives
\begin{equation}
b_{\eta^{\prime}}/\sqrt{N_c} =\langle 0| F \tilde{F}\left|\eta^{\prime}\right\rangle=\frac{8 \pi^2 N_c}{\lambda N_f}\langle 0| \partial_\mu J_A^\mu\left|\eta^{\prime}\right\rangle \,,
\end{equation}
with $J^\mu_A$ the axial current. Finally, using $\langle 0| \partial_\mu J_A^\mu\left|\eta^{\prime}\right\rangle =p^\mu\langle 0| J_A^\mu\left|\eta^{\prime}\right\rangle$ and 
\begin{equation}
\langle 0| J_A^\mu\left|\eta^{\prime}\right\rangle=-i \sqrt{N_f} f_\pi p_\mu
\end{equation}
we arrive at the Witten-Veneziano relation
\begin{align}
m_{\eta^{\prime}}^2=\frac{4 N_f}{f_\pi^2} \chi_\mathrm{YM}\,.
\label{eq:WV_app}
\end{align}
For a derivation of \eqref{eq:WV_main} including the dependence of $\chi_\mathrm{YM}$ on the $\eta$ and $K$ masses, see Ref.~\cite{Veneziano:1979ec}. Interestingly, since \eqref{eq:Lagrangian_largeN_YM} implies that the divergence of the axial current disappears in the large $N_c$ limit (as $1/N_c$), this relation suggests that $m_{\eta'}$ is not due to instantons, and instead comes from fluctuations in the topological charge density.

\section{Superposition of $\theta$ states and cluster decomposition principle}
\label{app:CDP}

Here we argue that the vacuum of QCD cannot be a superposition of $|\theta\rangle$ vacua, e.g, of the form
\begin{align}
    \ket{\alpha} = \frac{\ket{\theta_1} + \ket{\theta_2}}{\sqrt{2}}\,,
    \label{eq:vac_linear_combo}
\end{align}
with $\theta_1 \neq \theta_2$. It is not clearly physically meaningful  to write a superposition of states which live in different superselection sectors; doing so would require the Hilbert space to contain multiple $|\theta\rangle$ sectors. Suppose however that the Hilbert space does contain at least the sectors $\ket{\theta_1}$ and $\ket{\theta_2}$. In this case, we may first note that under large gauge transformation the state \eqref{eq:vac_linear_combo} does not transform by an overall phase, 
\begin{align}
    U\ket{\alpha} =\frac{e^{i\theta_1} \ket{\theta_1} + e^{i\theta_2} \ket{\theta_2}}{\sqrt{2}}\,,
\label{eq:vac_linear_combo_large_gauge_transformation}
\end{align}
so \eqref{eq:vac_linear_combo} breaks gauge invariance of the vacuum state. Though unusual, this is not so problematic as correlation functions of gauge-invariant operators $A$ are gauge-invariant; e.g, vacuum expectation values decompose as 
\begin{align}
\bra{\alpha}A\ket{\alpha} =\frac{1}{2}\bra{\theta_1}A\ket{\theta_1}+\frac{1}{2}\bra{\theta_2}A\ket{\theta_2}\,.
\label{eq:correlation_funcs}
\end{align}

What is more problematic is that  \eqref{eq:vac_linear_combo} being the vacuum is inconsistent with locality, and in particular the cluster decomposition property, as we now show. We denote the eigenstates of the Hamiltonian as $\ket{\theta}$. They form an orthonormal basis and therefore $\bra{\theta}\ket{\theta^{\prime}} = \delta(\theta-\theta^{\prime})$ and the identity decomposes as $\mathds{1} = \int_{\theta}\ket{\theta}\bra{\theta}$. To preserve the locality of the theory, the cluster decomposition principle tells us that given two operators $A(x)$ and $B(y)$ invariant under $SU(3)$ gauge transformations, with $x$ and $y$ space-like separated and $|x-y|\to\infty$,
\begin{align}
\begin{split}
    \bra{\alpha}A(x)B(y)\ket{\alpha} =& \bra{\alpha}A(x)\ket{\alpha} \bra{\alpha}B(y)\ket{\alpha}\\
    =&\frac{1}{4}\bigg(\bra{\theta_1}A(x)\ket{\theta_1} + \bra{\theta_2}A(x)\ket{\theta_2}\bigg)\\
    &\bigg(\bra{\theta_1}B(y)\ket{\theta_1} + \bra{\theta_2}B(y)\ket{\theta_2}\bigg)\,.
    \label{eq:D4}
\end{split}
\end{align}
Here we used that $\bra{\theta}A(x)\ket{\theta'}\propto\delta(\theta-\theta')$ for any gauge-invariant operator $A$. However, the relation above is clearly not satisfied by the state in \eqref{eq:vac_linear_combo}, as we obtain
\begin{align}
\begin{split}
    \bra{\alpha}A(x)B(y)\ket{\alpha} =& \int_{\theta}\bra{\alpha}A(x)\ket{\theta} \bra{\theta}B(y)\ket{\alpha}\\
    =&\frac{1}{2}\bra{\theta_1}A(x)\ket{\theta_1}\bra{\theta_1}B(y)\ket{\theta_1}\\
    +&\frac{1}{2}\bra{\theta_2}A(x)\ket{\theta_2}\bra{\theta_2}B(y)\ket{\theta_2}\,.
    \label{eq:D5}
\end{split}
\end{align}
As \eqref{eq:D4} and \eqref{eq:D5} differ, $\ket{\alpha}$ does not satisfy the cluster decomposition principle and is therefore not a valid vacuum state. 

\section{Gauging non-normal subgroups of a symmetry group}
\label{sec:gauge_non_normal_subgroups}

Here we justify that the $\mathbb{Z}_2$ subgroup, corresponding to spacetime parity, of the full Lorentz group in $d$ spacetime dimensions $O(1,d-1)$ cannot be gauged while preserving Lorentz symmetry unless the full $O(1,d-1)$ is gauged, as pointed out in Ref.~\cite{McNamara:2022lrw}. Note that this implies that gauging the 4D $P$ or $CP$  requires gravity (which gauges the Lorentz symmetry, as we explain in Section \ref{sec:gauging_CP_QG}).
In fact a more general claim holds \footnote{We thank Matthew Reece for discussions on this point.}: if $H$ is a non-normal subgroup of a symmetry group $G$ of a theory, then gauging $H$ is not possible without breaking the (global) symmetry $G$, unless $G$ is gauged.
\enlargethispage{\baselineskip}  

To see why this holds, note that a global symmetry transformation must send gauge-equivalent states to gauge-equivalent states. Let us consider the space $\mathcal{H}$ of states (where we have not yet declared states related by gauge transformations to be equivalent), and for brevity we will write operators $U(g)$ transforming in a representation of $G$ by the associated group element $g \in G$. For any $g \in G$, $h \in H$, and state $|\psi\rangle \in \mathcal{H}$, $g|\psi\rangle$ must be gauge equivalent to $gh|\psi\rangle$. That is, there exists some $h' \in H$ such that $h'g|\psi\rangle = gh|\psi\rangle$. 
As this reasoning must hold for any other arbitrary state $|\psi'\rangle$ as well as for the sum $|\psi\rangle + |\psi'\rangle$, $h'$ must the same for every state $|\psi\rangle$, so $h'g = gh$. Hence $H$ is stable under conjugation by $G$, meaning $H$ must be normal.

\end{document}